\pdfoutput=1

\documentclass[a4paper,fleqn]{cas-dc}

\usepackage[numbers,sort&compress]{natbib}   %

\usepackage[utf8]{inputenc}   %
\usepackage{siunitx}          %
\usepackage{subcaption}       %
\usepackage{booktabs}         %
\usepackage{tabularx}         %
\newcolumntype{P}[1]{>{\raggedright\arraybackslash}p{#1}}

\newcommand{\RQ}{\ensuremath{\mathrm{RQ}}}
\newcommand{\RQd}{\ensuremath{\mathrm{RQ}_{d}}}
\newcommand{\RQdx}{\ensuremath{\mathrm{RQ}_{dx}}}
\newcommand{\RQdy}{\ensuremath{\mathrm{RQ}_{dy}}}
\newcommand{\fy}{\ensuremath{f_{y}}}
\newcommand{\tlab}[1]{{\bfseries\boldmath #1}}   %

\begin{document}
\let\WriteBookmarks\relax
\def\floatpagepagefraction{1}
\def\textpagefraction{.001}

\shorttitle{A 3D workflow for replication quality and tool degradation in moulded microfluidic channels}
\shortauthors{M. O. Güngör et al.}

\title[mode=title]{A 3D measurement workflow for assessing replication quality and tool degradation in moulded microfluidic channels}

\author[1,2]{Mehmet Oğulcan Güngör}[orcid=0009-0007-8864-5012]
\credit{Conceptualization, Methodology, Software, Investigation, Data curation, Writing -- original draft}

\author[3,4]{Jérôme Charmet}[orcid=0000-0001-6992-4090]
\cormark[1]
\ead{jerome.charmet@unibe.ch}
\credit{Conceptualization, Writing -- review \& editing}

\author[1,2,5]{Ender Yıldırım}[orcid=0000-0002-7969-2243]
\cormark[1]
\ead{yender@metu.edu.tr}
\credit{Conceptualization, Supervision, Writing -- review \& editing}

\affiliation[1]{organization={Micro and Nanotechnology, Middle East Technical University},
            city={Ankara},
            postcode={06800},
            country={Türkiye}}

\affiliation[2]{organization={METU MEMS Center},
            city={Ankara},
            country={Türkiye}}

\affiliation[3]{organization={School of Engineering, HE-Arc Ingénierie, HES-SO University of Applied Sciences Western Switzerland},
            city={Neuchâtel},
            country={Switzerland}}

\affiliation[4]{organization={School of Biomedical and Precision Engineering, University of Bern},
            city={Bern},
            country={Switzerland}}

\affiliation[5]{organization={Department of Mechanical Engineering, Middle East Technical University},
            city={Ankara},
            postcode={06800},
            country={Türkiye}}

\cortext[1]{Corresponding authors}

\begin{abstract}
Mould-based replication, such as injection moulding with 3D-printed
inserts, needs a check of how faithfully microchannels are transferred from
master to polymer replica. It is usually limited to a width or depth ratio
from one cross-section, which misses the sidewall shape and depends on where
the section is taken. We present a workflow that assesses the whole channel
from 3D scans of master and replica, without sectioning. The scans are
registered on a common grid, and each height map is converted into a map of
local slope. The correlation of the two slope maps over the channel
interior, \RQd{}, is reported with the channel dimensions and height- and
slope-error maps. These outputs separate a change of shape from a change of
size and locate it. A user can thereby compare mould materials, detect tool
wear, and tell whether a deviation arose before or during moulding. On
synthetic channels with known degradation, \RQd{} fell steadily as the
sidewalls were rounded, whereas the height correlation levelled off. For
injection moulding with stereolithography-printed inserts in three resins,
\RQd{} ranged from $0.17$ to $0.91$ over fifteen insert-to-part pairs. The
height correlation placed the ten well-formed channels within $0.03$. Over a
twenty-shot run, one insert's \RQd{} fell from $0.90$ to $0.72$ as its
sidewalls wore, while its height correlation stayed above $0.93$.
We developed open-source software that implements the workflow. Because the
workflow needs only registered 3D scans of a master and its replica, it can
in principle be applied to other mould-based processes.
\end{abstract}


\maketitle
\flushbottom

\section{Introduction}\label{sec:intro}

Microfluidic devices are increasingly replicated from moulds made by
digital fabrication, such as 3D-printed or micromilled masters
\citep{naderi2019, charmet2020}. Mould-based processes need a metrological
check of fidelity to confirm that the microfeatures of the master are
transferred to the polymer replica. This need is independent of how the
master is produced. It has, however, grown more pressing as rapid tooling
expands the range of masters available for thermoplastic replication.
Photopolymer inserts printed by stereolithography (SLA), for example, can be
fabricated rapidly and used directly for injection moulding
\citep{rahmati2007, griffiths2011tooling, convery2021, menezes2024}. They
also serve as templates for thermoplastic embossing \citep{khoo2024} and as
masters for replica moulding in poly(dimethylsiloxane) (PDMS)
\citep{venzac2021, poskus2023}. In each case a quantitative assessment is
needed, not merely to establish whether the feature has formed, but to
determine how faithfully its spatial geometry has been reproduced.

Replication quality in mould-based microfabrication is conventionally
reduced to a single dimension, the transcription ratio: the height or width
of a feature on the part over that on the mould \citep{hansen2011,
loaldi2021}. A channel depth or width is measured on the insert and on the moulded part, from a section or a top-down image, and the two are compared with each other or with the nominal \citep{gulcur2021,
stampone2024, evens2022}. The practice runs through the microfluidic
moulding literature, from steel tools to printed ones. Demonstrations
typically report the replicated width and depth, sometimes with an average
roughness, and little else \citep{zhang2017, convery2021, menezes2024,
dempsey2020, ma2019, poskus2023}.
Such a ratio is easy to obtain but insensitive to several characteristics of the channel. Two channels with identical mean width can
differ completely in sidewall steepness, in the sharpness of the
floor-to-wall transition, and in surface texture. All of these can
influence flow, sealing, and optical access. The measurement is also operator-dependent: the result depends on where along the channel the section is taken, where the measurement cursors are placed and how the edge of a rounded, drafted wall is defined, so two users measuring the same part can report different values. The reference review of the
field lists the definition of edges, alignment, and the location of
measuring points among the practical challenges of this comparison
\citep{hansen2011}. A recent metrological treatment of the ratio reports a
pillar replication fidelity of $80.6\,\%$ or $96.2\,\%$, depending on which
histogram peak is taken as the replicated height \citep{loaldi2021}.
Finally, a single value from one section cannot reliably distinguish a
faithfully reproduced channel from one that has slumped, rounded, or only
partly filled. The problem is therefore one of measurement rather than of
process characterisation alone. What is needed is a defined measurement
procedure that assigns a comparable quantitative grade to the spatial
fidelity of a replicated feature. The methods reviewed next are judged
against that requirement.

Several richer comparison methods exist, each with limitations for microchannel assessment. Areal surface-texture parameters and
roughness-transfer studies quantify how texture statistics carry from tool to
part \citep{piccolo2021, vera2018, aboud2025}. The slope parameter $S_{dq}$
has been used beside $S_q$ because it responds to the morphology of a replicated texture, a height parameter only to its amplitude
\citep{baruffi2019}. But statistics carry no spatial correspondence: they
cannot say whether the steep and flat regions of two surfaces coincide. Two
surfaces can differ entirely in geometry yet share the same parameter values
\citep{vorburger2011, zheng2020}. A group that computed both found that
roughness parameters showed no clear tendency where a point-for-point
correlation did \citep{kumar2019}. Process-side practice adds the degree of
replication, the replicated feature height as a fraction of the master's
\citep{hansen2011, gulcur2021}, which grades filling but not shape. Areal part scans have also quantified warpage \citep{cetin2015}, a device-level form error, but not wall fidelity.

Correlation-based approaches do compare surfaces point for point.
Cross-correlation of registered topographies was developed at the National
Institute of Standards and Technology (NIST) for forensic toolmark comparison.
Vorburger et al.~\citep{vorburger2011, vorburger2015} defined the maximum
cross-correlation $\mathrm{CCF}_{\max}$ together with a normalised
difference $D_s$, and named the accuracy of surface replication among its
applications. Kumar et al.~\citep{kumar2019} applied both to moulded
replicas of natural surfaces, and correlation has since entered
manufacturing quality assurance \citep{zhang2020qa}. Perris et al.~\citep{perris2023} graded the replication of numerically generated rough
surfaces with a normalised cross-correlation ratio computed from line
profiles of the printed master and the moulded part. Zheng et al.~\citep{zheng2020} and Wang et al.~\citep{wang2021sim} correlated registered
height maps to decide whether two areal measurements portray the same
surface. Jiang et al.~\citep{jiang2021} replaced the height maps with their
Fourier amplitude spectra for the same purpose. These approaches correlate heights rather than gradients, and none targets the microchannel
problem. A height correlation is dominated by the gross channel form and
saturates once a channel has merely formed. The similarity frameworks decide whether two measurements represent the same surface rather than grade how faithfully one replicates another. Loaldi et~al.~\citep{loaldi2021} came closest to a gradient comparison. They
subtracted the slope maps of a diamond-machined insert and its replica along
one axis, and found a slope fidelity of $67\,\%$ where the height residual
gave $79\,\%$. The measure, however, is one-directional, its value depends
on a chosen uncertainty threshold, and it was applied once, to a steel tool.

For microchannels, an important step came from derivative correlation.
Working on ultrasonically embossed microchannels, Yildirim et al.~\citep{yildirim2024} cross-correlated the cross-sectional profiles of mould
and replica. They showed that correlating the derivatives of the profiles,
rather than the profiles themselves, is considerably more sensitive to
sidewall fidelity. In their data the profile (height) correlation stayed above $0.97$ in every case, shifting only in the third decimal even for a replica with a pronounced swallowtail defect. The derivative
correlation, by contrast, spread from $0.50$ to $1.00$ and separated good
from poor replication clearly. They adopted $0.8$ as a practical pass threshold. The gain comes from differentiation, which amplifies the
deviations at the sidewall, where the profile turns most steeply. The
method, however, inherits the limitations of its input. It is tied to a
single cross-section that must be obtained destructively, and one section
cannot represent along-channel variation. The choice of where to section
reintroduces the operator dependence of the dimensional ratio, and a
two-dimensional profile carries the derivative in only one direction.

To our knowledge, no metric used in replication metrology combines the four properties required here: full-length areal coverage of the
channel interior, sensitivity to local slope rather than height alone,
non-destructive acquisition, and evaluation without operator-chosen steps
after acquisition. Even existing areal correlation methods retain manual choices, such as the waviness cut-offs of Kumar et al.~\citep{kumar2019}
and the lateral magnification adjustment of the NIST comparisons
\citep{vorburger2011}. We therefore propose \RQd{} together with a complete
analysis workflow. The workflow registers 3D surface measurements of a master and its replica on a common grid. It converts each
height map into a map of local slope magnitude, $|\nabla Z|$, and
correlates the two slope maps over the shared channel interior. \RQd{} thus
grades the spatial correspondence of steep and flat regions over the full
channel length, including both sidewalls and the floor. It needs no destructive sectioning or manual choice of a cross-section. Unlike a
height correlation, it did not saturate at high fidelity in the tests
reported here. We further show that, for the channels studied, \RQd{} is
governed by the sidewalls. An open-source, in-browser tool implements the workflow and reports the conventional width, depth and wall angle, with spatial height- and slope-error maps, so the dimensional changes that a correlation intentionally discards remain available for interpretation.
Table~\ref{tab:rq_metrics} places the metric family in this landscape.

\begin{table*}[!tb]
\centering
\caption{Measures of microchannel replication quality reported in the
literature, and the $\mathrm{RQ}$ metric family introduced in this work.}
\label{tab:rq_metrics}
\footnotesize
\setlength{\tabcolsep}{4pt}
\renewcommand{\arraystretch}{1.15}
\begin{tabularx}{\textwidth}{@{}P{0.21\textwidth}P{0.215\textwidth}P{0.25\textwidth}>{\raggedright\arraybackslash}X@{}}
\toprule
Reported measure & Basis & What it captures & Limitations \\
\midrule
\tlab{Channel width and depth; ratio to nominal or master}
\citep{zhang2017, convery2021, menezes2024, dempsey2020, ma2019,
poskus2023} &
Few profiles, sections or top-down images; sometimes destructive &
Mean feature size; tolerance checks &
Single location; operator-dependent sectioning and edge definition; blind to
wall shape and texture \\[0.3ex] \addlinespace[0.45ex]
\tlab{Cross-section profile shape} (e.g.\ trapezoidal versus
hemispherical) \citep{stampone2024} &
Sectioned or scanned profiles &
Wall form, angle and corner radius; link to demoulding stress and tool
life &
Few locations; profile-bound; operator-chosen section \\[0.3ex] \addlinespace[0.45ex]
\tlab{Sidewall waviness and insert deflection} \citep{menezes2024} &
Optical inspection of the insert &
Tool-side degradation indicators &
Describes the tool, not master-to-replica correspondence \\[0.3ex] \addlinespace[0.45ex]
\tlab{Global form and warpage} \citep{cetin2015} &
Full-field 3D scan of the part &
Device-level deformation, e.g.\ for bonding &
Feature-level wall fidelity invisible \\[0.3ex] \addlinespace[0.45ex]
\tlab{Degree of replication}
\citep{hansen2011, gulcur2021, evens2022} &
Replicated over master feature height; process monitoring &
Filling completeness across process conditions &
Scalar filling measure; a filled channel scores well regardless of its
shape \\[0.3ex] \addlinespace[0.45ex]
\tlab{Roughness and texture transfer} ($S_a$, $S_q$, $S_{dq}$; texture
spectra) \citep{piccolo2021, vera2018, baruffi2019, aboud2025,
keane2025, tosello2026} &
Areal or AFM texture statistics, part-to-tool ratios and spectra &
Which texture scales transfer from tool to part; with $S_{dq}$, whether
slopes are reproduced on average &
No spatial correspondence; cannot localise or grade shape \\[0.3ex] \addlinespace[0.45ex]
\tlab{Areal height cross-correlation $\mathrm{CCF}_{\max}$ with normalised
difference $D_s$} \citep{vorburger2011, kumar2019} &
Pearson correlation of registered height maps at the in-phase shift; $D_s$
the mean-square residual over the master's mean square &
Replication fidelity, termed replication accuracy by its authors (0.76--0.98 for moulded replicas of natural surfaces); $D_s$
restores amplitude sensitivity &
Height-based, dominated by gross form; $\mathrm{CCF}_{\max}$ blind to
vertical scale; manual filter cut-offs \\[0.3ex] \addlinespace[0.45ex]
\tlab{Normalised cross-correlation ratio} \citep{perris2023} &
Line profiles of master and replica (stochastic rough surfaces) &
Replication grading of transferred texture (up to ${\sim}$0.74 for
injection-moulded polystyrene); flat over a 30-shot run &
Height-based; not wall-sensitive \\[0.3ex] \addlinespace[0.45ex]
\tlab{Profile height correlation $r$} \citep{yildirim2024} &
Pearson correlation of one sectioned profile pair &
Gross profile correspondence &
Saturates: above 0.97 throughout, third-decimal shifts even at a gross
defect \\[0.3ex] \addlinespace[0.45ex]
\tlab{Profile derivative correlation $r_d$} \citep{yildirim2024} &
Pearson correlation of the differentiated profiles &
Sidewall fidelity along one section (0.50--1.00 spread; 0.8
threshold) &
Destructive single section; one derivative direction; operator-chosen
location \\[0.3ex] \addlinespace[0.45ex]
\tlab{Height and slope residual maps} \citep{loaldi2021} &
Registered insert-minus-replica maps &
Where and by how much the replica departs; fidelity (share of pixels within
measurement uncertainty) stricter for slope than height (67 versus
79\,\%) &
One derivative direction; value set by the uncertainty threshold; a
residual, not a bounded similarity \\[0.3ex] \addlinespace[0.45ex]
\tlab{Surface-matching correlation}
\citep{zheng2020, wang2021sim, jiang2021} &
Pearson correlation of registered height maps, or of their 2D Fourier
amplitude spectra &
Whether two measurements represent the same surface &
Built for identification, not grading; the spectral variant discards
spatial phase; cannot localise \\[0.3ex] \addlinespace[0.45ex]
\tlab{Image-quality indices}: structural similarity (SSIM) and gradient-magnitude
similarity deviation (GMSD) \citep{wang2004ssim, xue2014gmsd} &
Windowed height statistics (SSIM); pointwise gradient-magnitude similarity
pooled by its deviation (GMSD) &
Perceptual structural similarity of intensity images (SSIM); spatial
non-uniformity of the gradient distortion (GMSD) &
Built for perceived image quality, not metric height fields; SSIM inherits
height saturation; GMSD pooling misses uniform degradation \\
\midrule[0.8pt]
\addlinespace[0.3ex]
\multicolumn{4}{@{}l}{\emph{This work: the $\mathrm{RQ}$ metric family
(non-destructive, full-surface, deterministic from the scan)}} \\[0.3ex] \addlinespace[0.45ex]
\tlab{$\mathrm{RQ}$} &
Areal correlation of the height fields $Z$ &
Gross height correspondence over the full surface &
Floor and field dominate; saturates once the channel forms \\[0.3ex] \addlinespace[0.45ex]
\tlab{$\mathrm{RQ}_{dx}$} &
Areal correlation of $\partial Z/\partial x$ &
Along-channel slopes, longitudinal features, transferred floor texture &
Absolute value low when fine texture is not shared; relative trend more
informative \\[0.3ex] \addlinespace[0.45ex]
\tlab{$\mathrm{RQ}_{dy}$} &
Areal correlation of $\partial Z/\partial y$ &
Across-channel slopes and sidewall form &
Direction-specific; depends on consistent channel alignment \\[0.3ex] \addlinespace[0.45ex]
\tlab{$\mathrm{RQ}_{d}$ (primary)} &
Areal correlation of $|\nabla Z|$ &
Full-surface spatial correspondence of local steepness; sidewall-dominated;
remains discriminating where the height correlation saturates &
Discards gradient sign and direction; invariant to positive amplitude
scaling; may dilute very localised defects ($E_Z$, $E_d$ complement) \\
\bottomrule
\end{tabularx}
\end{table*}

The metric is first characterised on controlled synthetic surfaces with known ground truth, where its response to wall degradation, depth loss,
noise, local defects and registration can be separated systematically. It is then demonstrated on SLA rapid tooling for injection-moulded microfluidics. Three commercial insert
resins and five channel widths provide fifteen master-to-replica
comparisons spanning good to failed replication. Repeated scans over a
twenty-shot moulding run test whether the workflow tracks progressive tool degradation. The same measurements locate fidelity loss along the
CAD-to-insert-to-part chain and quantify the dimensional and surface changes. \RQd{} requires only
registered height maps of a master and its replica, and can in principle be applied to other mould-based processes.

In practice the workflow supports five recurring tasks: characterising a
single channel, assessing an insert against its design, assessing a moulded
part against its insert, following an insert over a moulding run, and
assessing the whole chain from design to part. Within a replication
comparison, it separates deviations of shape from deviations of size and
locates them on the channel. On the injection-moulded channels
studied here, the slope correlation separated well-formed replicas that a
height correlation rated alike. It also followed the progressive wear of an
insert that the height correlation barely registered.

\section{Materials and methods}\label{sec:methods}

\subsection{Replication-quality metrics}\label{sec:methods-metric}

The purpose of \RQd{} is to answer a spatial question: after two channel
surfaces have been aligned, do their steep and flat regions occur at
corresponding locations? In a faithfully replicated channel, high-gradient
regions occur at matching sidewalls and floor-to-wall transitions, while the
channel floor and the surrounding field remain comparatively flat. Sidewall
rounding, lateral displacement, surface collapse, and spatially distributed
defects all alter the position and distribution of these gradients. \RQd{}
captures such changes by converting each measured surface into a
gradient-magnitude map and correlating the two maps cell by cell.

\begin{figure}[pos=!tb]
\centering
\includegraphics[width=\columnwidth]{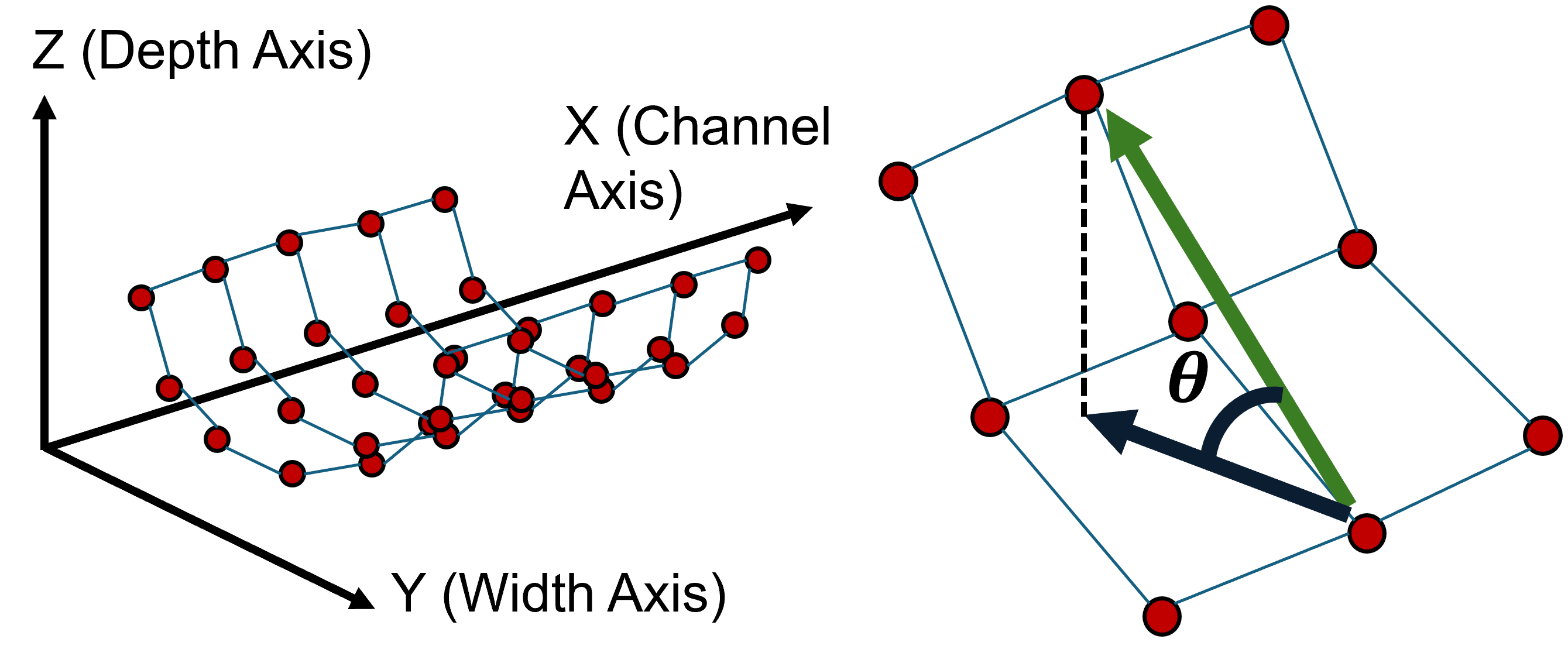}
\caption{Coordinate convention (left) and estimation of one local
directional-gradient component from neighbouring height values on the
analysis grid (right).}
\label{fig:local_gradient}
\end{figure}

Each compared surface is represented as a single-valued height field
$Z(x,y)$, with $x$ the along-channel and $y$ the across-channel coordinate
(Fig.~\ref{fig:local_gradient}).
Let $Z_A$ and $Z_B$ denote the two co-registered height fields of the
reference surface~A (the design, the insert, or the insert in an earlier
state) and the surface~B compared with it, sampled on a common regular grid
of cell size $\Delta$ spanning their overlapping extent. A
common valid-cell mask excludes any cell missing on either surface, for
example through optical dropout, so every correlation is evaluated only over
cells valid on both. Both conventions follow the NIST topography
comparisons. Their $\mathrm{CCF}_{\max}$ is defined at the in-phase shift,
since a correlation between misaligned topographies is meaningless
\citep{vorburger2015, kumar2019}. Their cell-wise correlation was introduced
so that invalid regions could be kept out of the score \citep{vorburger2015}.
The valid cells are further restricted to the channel interior, the cells
that lie deeper than a fixed fraction of the channel depth on both surfaces,
so that the flat field around the channel does not dominate the score. Height
gradients are computed by second-order central differences,
\begin{equation}\label{eq:centdiff}
  \frac{\partial Z}{\partial x}\bigg|_{i,j} \approx
  \frac{Z_{i+1,j}-Z_{i-1,j}}{2\,\Delta},
\end{equation}
and analogously in $y$, on an isotropic grid of cell size $\Delta$.
Figure~\ref{fig:local_gradient} illustrates the geometric meaning
of one directional component. The local height difference between
neighbouring cells is the surface slope in that direction. For a locally
planar wall, it relates to the local wall angle through
$|\partial Z/\partial y| \approx \tan\theta$. Because the channel is aligned
with the analysis axes, the $x$ component describes variation along the
channel while the $y$ component cuts through the sidewalls.

The two directional components are combined into the scalar gradient
magnitude
\begin{equation}\label{eq:grad}
  G(x,y) = |\nabla Z(x,y)| = \sqrt{\left(\frac{\partial Z}{\partial
  x}\right)^{2} + \left(\frac{\partial Z}{\partial y}\right)^{2}},
\end{equation}
a direction-independent measure of local steepness. The magnitude prevents
cancellation between positive and negative wall slopes. It lets sidewalls,
floor-to-wall transitions, and longitudinal defects all contribute to one
full-surface map. Each height map is thereby reduced to a gradient-magnitude
map, and \RQd{} is the Pearson correlation \citep{pearson1895} of the two
magnitude maps. The gradient vectors themselves are never correlated
directly. Writing $G_{A,k}$ and $G_{B,k}$ for the magnitudes at the $k$-th
of the $N$ cells in the mask,
\begin{equation}\label{eq:rqd}
  \RQd =
  \frac{\sum_{k}\bigl(G_{A,k}-\overline{G}_{A}\bigr)\bigl(G_{B,k}-\overline{G}_{B}\bigr)}
  {\sqrt{\sum_{k}\bigl(G_{A,k}-\overline{G}_{A}\bigr)^{2}}\,
   \sqrt{\sum_{k}\bigl(G_{B,k}-\overline{G}_{B}\bigr)^{2}}},
\end{equation}
where $\overline{G}_A$ and $\overline{G}_B$ are the mean magnitudes over
the mask.

Because the gradient is a finite difference on the analysis grid, the
metric is defined at a stated cell size. Three operations scale with the
same $\Delta$. The measured points are bin-averaged onto cells of side
$\Delta$. A NaN-aware Gaussian filter of $\sigma = \Delta$ then regularises
the field before differentiation and fills an empty cell from measured cells
up to four cells away. Finally, the central difference spans $2\Delta$. We therefore write $\RQd(\Delta)$ where the dependence matters,
in the way a slope parameter is reported together with its nesting index
\citep{leach2013}. Values at different cell sizes are not comparable. The
cell size follows from the instrument, not from the
data. A square cell of side $\Delta \geq \sqrt{2}\,s$ contains at least
one point of a lattice of pitch $s$, whatever the in-plane rotation of the
alignment. $\Delta$ is therefore taken as the smallest round value
satisfying $\Delta \geq \sqrt{2}\,s$, with $s$ the finest lateral
resolution of the instrument. Every cell is then a measured average rather
than an interpolation.

The related metrics are defined on the same grid and mask. The
directional correlations
\begin{align}\label{eq:rqdxy}
  \RQdx &= \operatorname{corr}\bigl(\partial Z_A/\partial x,\; \partial Z_B/\partial x\bigr), \nonumber\\
  \RQdy &= \operatorname{corr}\bigl(\partial Z_A/\partial y,\; \partial Z_B/\partial y\bigr)
\end{align}
correlate the signed directional derivatives. The height correlation
$\RQ = \operatorname{corr}(Z_A,\, Z_B)$ correlates the height fields. Each
is evaluated with the Pearson coefficient of Eq.~\eqref{eq:rqd} applied to
the indicated quantity. Two spatially resolved
error maps accompany the scalars. Error here means the replication
deviation of one surface from the other, not a measurement error. The maps
retain the absolute-deviation information that the correlations discard.
They are
the signed height difference and the unsigned gradient-magnitude
difference,
\begin{equation}\label{eq:errmaps}
  E_Z = Z_A - Z_B, \qquad
  E_d = \bigl|\,|\nabla Z_B| - |\nabla Z_A|\,\bigr|,
\end{equation}
where $E_Z$ shows where one surface is higher or lower than the other and
$E_d$ highlights poor shape fidelity such as rounded or displaced sidewalls.

This scheme generalises the two-dimensional derivative cross-correlation of Yildirim et al.~\citep{yildirim2024}. \RQ{} is the
areal counterpart of their profile correlation. \RQdx{}, \RQdy{} and \RQd{}
are the areal counterparts of their derivative correlation, resolved into both directions and evaluated over the whole surface instead of one destructively obtained section.

\subsection{Analysis software}\label{sec:software}

\begin{figure*}[!tb]
\centering
\includegraphics[width=\textwidth]{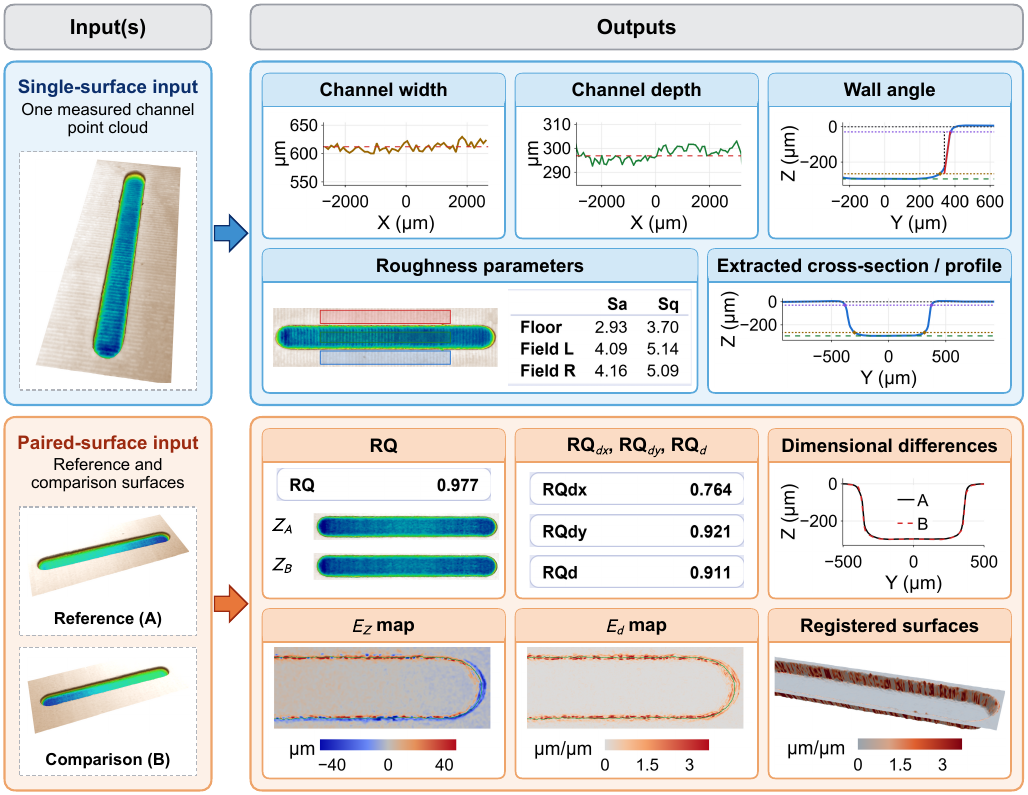}
\caption{Inputs and outputs of the analysis software, all panels rendered by
it. Top: single-surface analysis of one channel scan (Rigid 10K
\SI{600}{\micro\meter} insert). Bottom: paired comparison of a reference
surface A (High Temp \SI{600}{\micro\meter} insert) with a comparison surface B
(its first moulded part); the registered surfaces are coloured by $E_d$.}
\label{fig:workflow}
\end{figure*}

The measurements are processed, and the metrics are computed, by the
Microchannel and Replication Quality Analyzer (MaRQ, version 1.0). This
open-source analysis pipeline, developed in this work, operates on
three-di\-men\-sion\-al point clouds from optical profilometry
(Fig.~\ref{fig:workflow}). A single-surface workflow extracts the geometry
and roughness of one surface, and a paired-surface workflow compares two
registered surfaces.

\begin{table}[pos=!tb]
\centering
\caption{Principal analysis parameters of the software, fixed for every
result in this paper.}
\label{tab:params}
\begin{tabularx}{\linewidth}{@{}P{0.32\linewidth}>{\raggedright\arraybackslash}X@{}}
\toprule
Parameter & Value \\
\midrule
Voxel size & \SI{5}{\micro\meter} \\
Aggregate cross-section & median per bin, 400 transverse bins over the channel length \\
Floor level; depth & median of profile bins $\leq$ 5th percentile; depth $d$ from the field to the floor level \\
Grid lines & \SI{5}{\micro\meter} cells, \SI{10}{\micro\meter} apart, central 80\,\% of the length; median \\
Width levels & floor 0.90, mid 0.50, opening 0.10 of the depth \\
Wall angle & line fitted at 0.25 to 0.75 of the depth; $<3^{\circ}$ not resolved \\
Longitudinal spread & SD over the grid lines; depth: 80 slices, SD over the central 80\,\% \\
Floor roughness strip & $0.5\times$ nominal width $\times$ \SI{4000}{\micro\meter}, points deeper than $0.7\,d$ \\
Field roughness strips & 400 $\times$ 4000~\si{\micro\meter}, symmetric about the axis, points above $-0.3\,d$ \\
Roughness form removal & least-squares plane per region, no further filtering \\
Correlation grid cell size $\Delta$ & \SI{5}{\micro\meter} ($\sqrt{2} \times$ \SI{3.51}{\micro\meter} lateral resolution, rounded) \\
Pre-gradient filter & NaN-aware Gaussian, $\sigma = \Delta$ \\
Channel-interior mask & cells below 0.20 of the depth on both surfaces \\
\bottomrule
\end{tabularx}
\end{table}

Each point cloud, in delimited text or \mbox{ISO~25178-72} X3P format, is
cleared of optical outliers and thinned to \SI{5}{\micro\meter} voxels. It is then
levelled on the flat field around the channel, which becomes the datum
$z = 0$. A raised feature, such as the channel-forming ridge of an insert,
is turned into the equivalent depression by a $180^{\circ}$ rotation about
an in-plane axis, so that it can be compared directly with the channel it
moulds. The channel axis is aligned with $x$ and the channel centred at the
origin, so that $x$ runs along the channel and $y$ across it. The channel depth is read from an aggregate
cross-section of the whole channel length. The widths, the wall angle and the
rounding of the floor and rim corners are read on grid lines across the
channel, spread over the central $80\,\%$ of its length. On each line the wall
lies where the height profile crosses a fixed fraction of the depth, located
by interpolation between neighbouring grid points, and each reported value is
the median over the lines. The wall angle is the inclination, from the
vertical, of a straight line through the wall positions between $25$ and
$75\,\%$ of the depth; the corner rounding is the distance by which the wall
departs from that line at the floor and opening levels. Walls within
$3^{\circ}$ of the vertical span too few grid cells to be resolved and are
reported as $<3^{\circ}$. The
variation along the channel is the spread over the lines, and over $80$
slices for the depth. Areal roughness (ISO~25178-2) is evaluated on a floor
strip and two field strips placed automatically from the detected channel.%
\looseness=-1

\begin{figure*}[!tb]
\centering
\includegraphics[width=\textwidth]{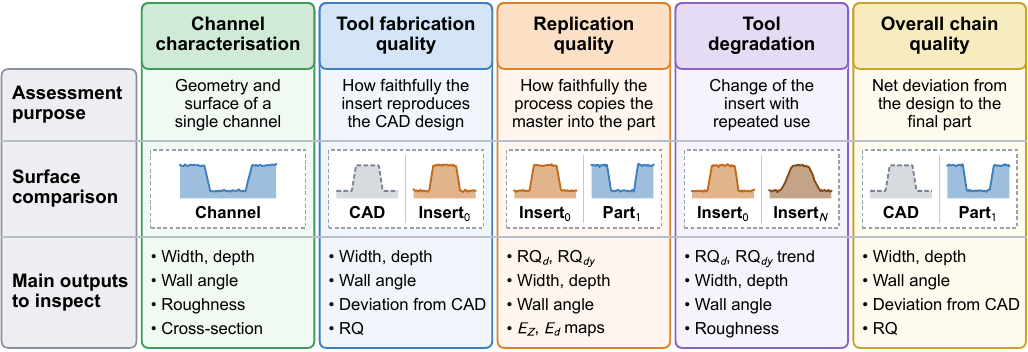}
\caption{Assessment tasks addressed by the workflow. CAD denotes the design
geometry, and subscripts give the shot number: Insert$_0$
is the as-printed insert, Insert$_N$ the same insert after $N$ shots and
Part$_1$ the first moulded part.}
\label{fig:decision}
\vspace{-2pt}
\end{figure*}

For a paired comparison, both surfaces are thus already in the same channel
frame, so their registration needs no pairwise search. They are brought onto
a common grid with a shared valid-cell mask, and the correlations and error
maps are evaluated over the channel interior. The software also resolves the
$180^{\circ}$ orientation ambiguity of a straight channel, a correction
required for $12$ of the $15$ insert-to-part comparisons.

Every step is deterministic with the fixed parameters of
Table~\ref{tab:params}, so repeated analyses of the same scan return
identical numbers. The extraction replaces manual cursor placement and is
operator-independent after acquisition. Sample placement, the scanned region
and the instrument settings remain acquisition decisions, held fixed here.
The software runs in a web browser without installation; its interface is
shown in Fig.~S1.1, and the complete algorithm and its validation are given
in Section~S1 of the supplementary material, with all parameters in
Table~S1.1.

\subsection{Interpretation and assessment tasks}\label{sec:methods-interp}

\RQd{} is a dimensionless measure of spatial correspondence, not a
percentage: $\RQd = 0.8$ does not mean that 80\,\% of the geometry was
replicated. The Pearson coefficient is bounded between $-1$ and $1$. In
correctly registered, physically representative channel comparisons it
principally occupies the interval from about $0$ (no shared gradient
structure) to $1$ (perfect linear correspondence of the steepness pattern).
Small negative values can arise from finite texture realisations, substantial ones only under severe misregistration.
The Pearson coefficient is invariant to constant offset and positive
scaling. A high \RQd{} therefore does not by itself establish identical
dimensions: two surfaces can share the spatial pattern of steep and flat
regions while differing in width, depth, or absolute wall angle. The metric is therefore
read together with the outputs that retain what it discards: the extracted
dimensions for size, $E_Z$ for height deviations, and $E_d$ for slope
deviations.
Interpreted directionally, \RQdy{} reflects the correspondence of the
across-channel walls and is the principal contributor to \RQd{} for channel
geometries. \RQdx{} reflects longitudinal slope and texture correspondence.
Its absolute value can stay low when fine floor texture is not shared
between independent scans, so its relative trend is more informative than
its level. We further quantify wall dominance with the ratio
$\fy = \sum_{k} (\partial Z/\partial y)_{k}^{2} / \sum_{k} |\nabla Z|_{k}^{2}$,
the share of the squared slope magnitude carried by the across-channel
component. The software computes it for each analysed surface over the
channel interior (cells deeper than $0.20$ of the depth) along the central
$60\,\%$ of the crop, where the rounded channel ends do not contribute. For
the insert-to-part pairs it is that of the part. It shows that, for the
channels studied, \RQd{} behaves as a measure of sidewall fidelity.\looseness=-1

Which surfaces are compared, and which of these outputs are read, depends
on the question (Fig.~\ref{fig:decision}). A single channel is characterised
by its dimensions and roughness alone. The fabrication of a mould insert is
assessed against its CAD design, where the height correlation and the
dimensions are read. Against the idealised CAD surface, \RQ{} does not
saturate, and the difference is dominated by systematic wall offsets that
the extracted dimensions measure directly. Replication is assessed
by comparing the insert with its moulded part. Here \RQd{} and \RQdy{} grade
the shape, the dimensions its size, and $E_Z$ and $E_d$ locate the
deviations. \RQd{} is read against the level that the same surface reaches
against its own rescan. Tool degradation
is followed by comparing an insert with its own initial state over a moulding
run, and the whole chain by comparing the moulded part with the design.
Printing deviations already appear in the insert against its design, whereas
moulding deviations appear only in the part against its insert. The workflow sets no acceptance thresholds. Tolerances depend on the device and
its function, and a user sets them from the application, for example from a
reference pair of known quality.

\begin{figure*}[!tb]
\centering
\includegraphics[width=\textwidth]{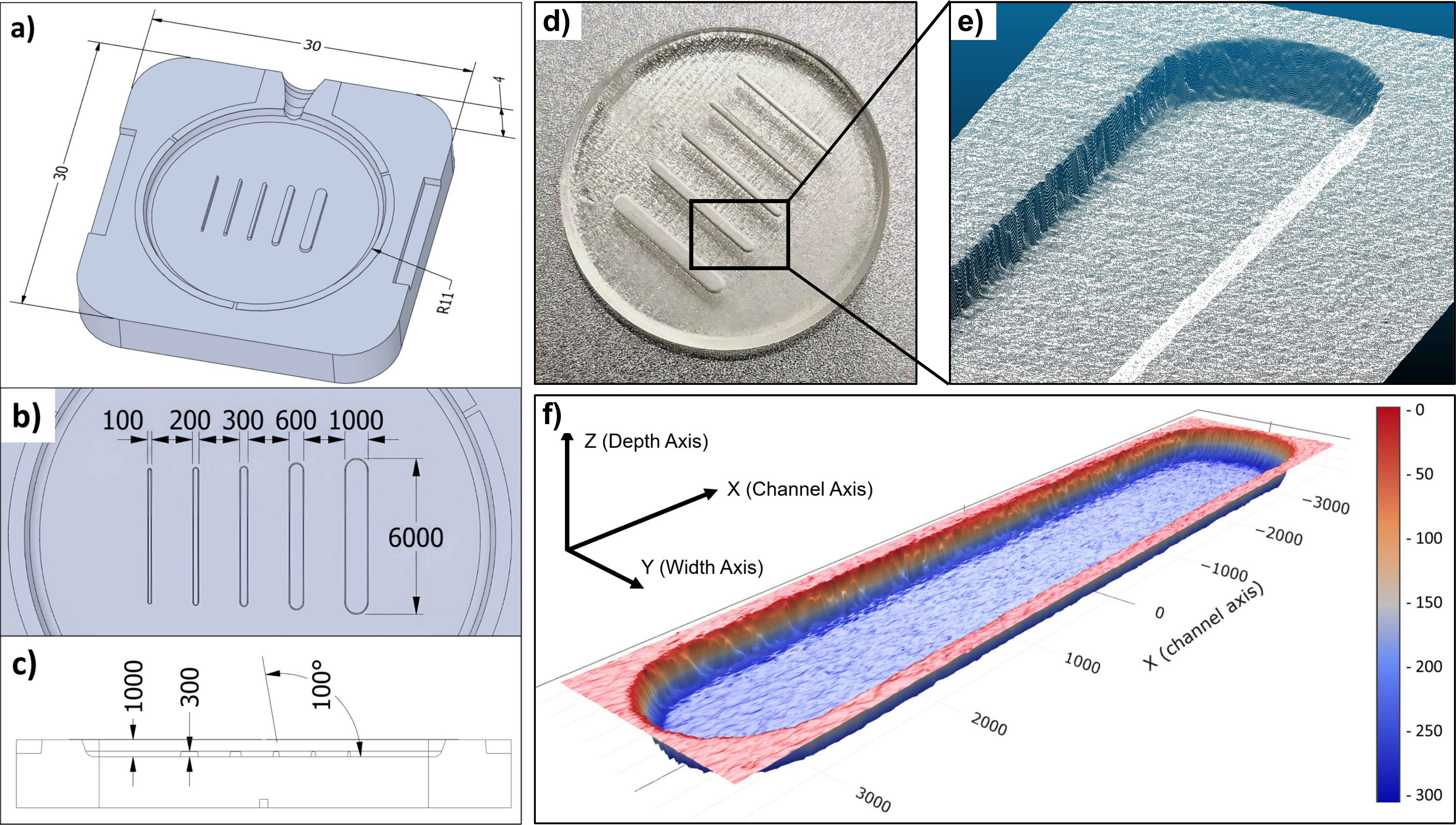}
\caption{Test geometry and measurement. (a)~Mould insert (mm); (b)~plan view
of the five channel positives; (c)~cross-section of the insert. (d)~Moulded
part; (e)~raw point cloud of the \SI{1000}{\micro\meter} channel; (f)~its
height field in the software (µm).}
\label{fig:insert_design}
\end{figure*}

\subsection{Synthetic test surfaces}\label{sec:methods-synthetic}

The metric was characterised on synthetic height fields with known ground
truth, built with the design geometry of the \SI{600}{\micro\meter} test
channel. The reference is a
trapezoidal channel \SI{300}{\micro\meter} deep with a
\SI{600}{\micro\meter} floor and a $10^{\circ}$ draft, on a $4 \times 2$~mm
field sampled at \SI{5}{\micro\meter}. A band-limited random microtexture of
\SI{3}{\micro\meter} standard deviation and about \SI{12}{\micro\meter}
correlation length is added to floor and field. A degraded copy is compared with
the undegraded reference through the production pipeline, with the same fixed
settings as the measured data. Four degradations are applied in controlled
amounts. Wall smoothing is a Gaussian rounding of the cross-section profile
of scale $\sigma$ that leaves the microtexture intact. Global smoothing is an
isotropic Gaussian blur of the whole field. Depth fill reduces the depth at
fixed opening width, which lowers the wall slope in proportion. The fourth
is added Gaussian noise. A localised defect is a Gaussian feature of
\SI{90}{\micro\meter} height and \SI{140}{\micro\meter} width placed on
the channel rim. Two reference cases bracket the scale: the reference compared
with itself, and two fields with the channel removed and independent
textures, evaluated over $30$ texture seeds. Registration bias is the residual
systematic change in \RQd{} of a re-registered copy relative to the aligned
copy. It was assessed by imposing rigid in-plane translations and rotations
and re-registering with the production pipeline. It is not corrected for,
and its maximum over the transforms is quoted. The seed-to-seed standard
deviation was obtained by repeating the degradation series with different
texture seeds.

\subsection{Moulding experiment}\label{sec:methods-experiment}

The workflow was applied to microchannels injection moulded with
stereolithography-printed inserts.

\subsubsection{Test geometry and insert fabrication}\label{sec:methods-materials}\label{sec:methods-fab}

The mould consists of two mating insert halves (Fig.~\ref{fig:insert_design}).
One half carries the microchannel positives, the raised ridges that mould the
channels, on its cavity floor. The other half is identical but smooth, so the
moulded part receives channel features on one face only. The patterned half
carries five straight microchannel positives. All five are
\SI{300}{\micro\meter} tall and \SI{6}{\milli\meter} long and span five widths,
$100$, $200$, $300$, $600$, and \SI{1000}{\micro\meter}, giving depth-to-width
aspect ratios of $3.0$, $1.5$, $1.0$, $0.5$, and $0.3$. This fixed-height,
variable-width set exposes the replication process to a broad range of feature
aspect ratios within a single insert. A $10^{\circ}$ draft angle on the sidewalls
eases release of the moulded part. The halves were designed in Autodesk Inventor
and exported as STL files. The same CAD surface serves as the reference for
the CAD-to-insert comparison, so the printed geometry is graded against the
design it was sliced from. On these CAD references the software's readings
return the design widths within \SI{0.1}{\micro\meter} and the $10^{\circ}$
wall angle within $0.02^{\circ}$, a check of the extraction on known
geometry (Section~S1 of the supplementary material).
The inserts were printed in three commercial Formlabs photopolymers, chosen to
span a wide range of moulding behaviour: White, Rigid~10K (R10K), and
High~Temp (HT).%
\looseness=-1

Inserts were printed on a Form~3+ low-force stereolithography (SLA) printer
(Formlabs, Somerville, MA, USA), with orientation, support, and slicing
prepared in Formlabs PreForm.
Every insert was built in the same orientation, tilted $30^{\circ}$ to the
platform with the microchannel cavity face upward.
The layer thickness was set to the finest available for each resin,
\SI{25}{\micro\meter} for HT and \SI{50}{\micro\meter} for R10K and White.
The remaining low-level parameters are fixed for each validated resin.
After printing, the inserts were
washed in isopropyl alcohol and UV post-cured to the manufacturer's
specification. The support structures were then removed and the back face was
sanded through a $320$, $600$ and $1500$~grit sequence to clear the support
witness marks and let the insert seat squarely in the steel holder. The microchannel cavity face,
never in contact with supports, required no such treatment. A final
isopropyl-alcohol wash removed the sanding dust, and each insert was inspected
under an optical microscope for defects.

\subsubsection{Injection moulding and the moulding run}\label{sec:methods-moulding}

Parts were injection moulded in polystyrene on a desktop injection moulding
machine (MSE Teknoloji, Gebze, Türkiye). The full study is a matrix of three resins $\times$
five channel widths. It yields the fifteen insert-to-part comparisons
analysed below, each between an insert and its first moulded part. To follow
the tool over the run, each insert was scanned in its as-fabricated state
(shot~0) and again after shots $1$, $5$, $10$, and $20$ where it survived.
Each rescan is compared with the insert's own shot-0 state, a fixed reference
that isolates the degradation of the tool.

\subsubsection{Optical profilometry}\label{sec:methods-metrology}

Surfaces were measured on an InfiniteFocus~G5 optical 3D measurement system
(Bruker Alicona, Graz, Austria), which reconstructs topography by focus
variation \citep{danzl2011}. The $5\times$ objective (numerical aperture
$0.15$) was used, with the instrument's vertical-resolution setting at
\SI{500}{\nano\meter}. The
objective's finest lateral
topographic resolution is \SI{3.51}{\micro\meter} \citep{alicona_g5}, and
the exported point clouds are sampled at that pitch. With
$s = \SI{3.51}{\micro\meter}$, the rule $\Delta \geq \sqrt{2}\,s$ gives
$\sqrt{2}\,s = \SI{4.96}{\micro\meter}$, rounded to
$\Delta = \SI{5}{\micro\meter}$. Every value in this paper is
$\RQd(\SI{5}{\micro\meter})$, written as \RQd{} for brevity.
Each scan covered at least one full channel feature, including its floor, both
sidewalls, and the adjacent field, producing dense three-dimensional point
clouds. Figure~\ref{fig:insert_design}(d)--(f) shows a moulded part, the raw
point cloud of one of its channels and the height field the software builds
from it, with the coordinate convention used throughout. Areal
surface-texture parameters follow ISO~25178 \citep{leach2013}. Registration
of each surface pair to a common frame and grid is performed by the analysis
software. No destructive preparation is
required.

\subsection{Reference similarity measures}\label{sec:methods-benchmark}

For comparison, three established similarity measures are evaluated on the same registered grids and channel-interior mask. The NIST
normalised difference \citep{vorburger2011, walton2016},
$D_s = R_q^2(Z_A - Z_B)/R_q^2(Z_A)$, is the mean-square roughness of the
difference topography over that of the reference. It is reported in the
similarity form $1 - D_s$ used by Kumar et~al.~\citep{kumar2019} and is
unbounded below. The structural similarity
index (SSIM) \citep{wang2004ssim} is a windowed comparison of the local
means, variances and covariance of two images. It is computed on the height
maps with its standard $11 \times 11$ Gaussian window ($\sigma = 1.5$
cells), $K_1 = 0.01$ and $K_2 = 0.03$, and the dynamic range set to the joint height range of the pair, and pooled by its mean. The gradient-magnitude similarity of Xue
et~al.~\citep{xue2014gmsd} compares the two gradient magnitudes cell by cell,
with a stabilising constant $c = 0.0026$ on heights scaled jointly to
$[0, 1]$ (Section~S7 of the supplementary material). It is pooled in two ways: by its standard deviation, the gradient-magnitude similarity deviation (GMSD), a distance that vanishes for identical maps, and by its mean, the gradient-magnitude similarity mean (GMSM) of the same authors. Its
magnitudes come from the same central differences as \RQd{}, not from the
Prewitt operator of the image implementation, so all measures share one
regularisation; no image down-sampling is applied. The SSIM and
gradient-similarity maps are computed over the whole valid crop and pooled
over the channel-interior mask. Differences between the measures thus
reflect their definitions rather than the cells they are evaluated on. The
six values of a pair (\RQ{}, \RQd{}, $D_s$, SSIM, GMSD and GMSM) come from
one registration and one grid, and are listed per comparison in Tables~S7.1 to S7.3.

\begin{figure*}[!tb]
\centering
\includegraphics[width=\textwidth]{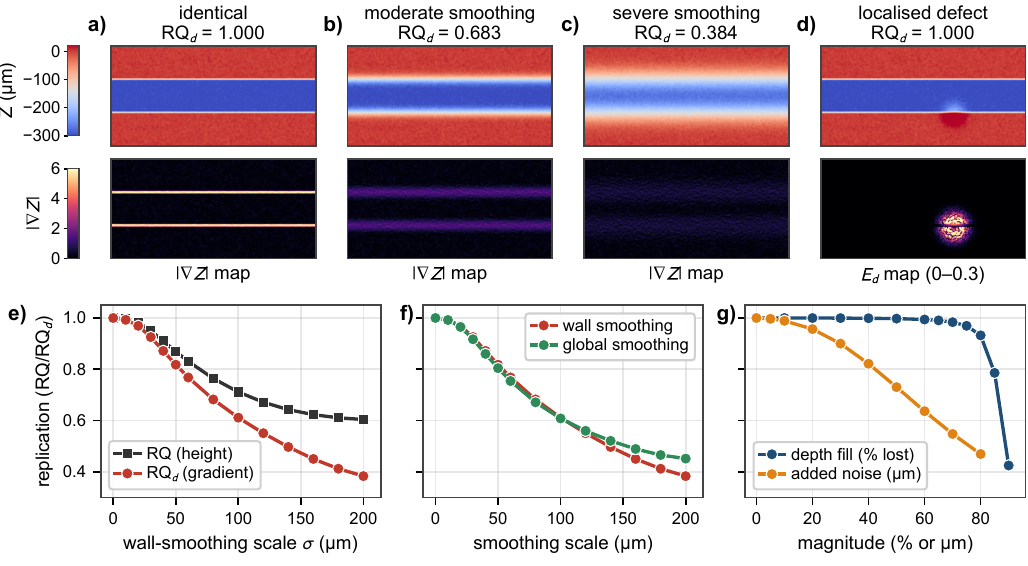}
\caption{Synthetic characterisation of $\mathrm{RQ}_{d}$.
(a)--(d)~Reference cases, with the height field $Z$ above and $|\nabla Z|$
below, scales on the left; in (d) the lower map is the gradient error $E_d$
on its own scale.
(e)~$\mathrm{RQ}$ and $\mathrm{RQ}_{d}$ against the wall-smoothing scale
$\sigma$. (f)~$\mathrm{RQ}_{d}$ against wall and global smoothing.
(g)~$\mathrm{RQ}_{d}$ against depth fill and added noise; (f) and (g) share
the y axis of (e).}
\label{fig:synthetic}
\vspace{4pt}
\end{figure*}

\subsection{Analysis settings and sensitivity protocol}\label{sec:methods-settings}

Spatial sampling is the principal analysis parameter, because it sets the
scale at which every downstream quantity is evaluated. It enters twice. Each
cloud is first thinned to \SI{5}{\micro\meter} voxels before gridding
(Table~\ref{tab:params}), a concession to in-browser processing. At this
voxel size the software recovers the widths of synthetic channels of known
geometry within a few micrometres and their wall angles within $0.5^{\circ}$
(Section~S1 of the supplementary material). The correlations depend, in
addition, on the grid cell size at which the gradients are taken. The cell
size $\Delta = \SI{5}{\micro\meter}$ is fixed from the instrument's lateral
resolution and not from the results, and is stated with every value in this
paper. The thinning leaves about $7\,\%$ of these cells to the filter,
against almost none with native-density clouds. Relative to gridding the
native points, it raises \RQd{} by less than $0.01$ on average and by at
most $0.02$. The thinning is therefore part of the stated definition.

The sensitivity of the correlations to the analysis settings is quantified
on the fifteen insert-to-part pairs, with crops and registration unchanged.
Four settings are varied in turn about their defaults: the grid cell size,
the channel-interior fraction ($0.20$ of the depth), the pre-gradient filter
and a rigid shift across the channel. The grid cell size is also varied on
the tool-degradation trajectories. As a robustness check, with all other
settings fixed, the default channel-interior mask (both surfaces below the
threshold) is compared with masks defined by the reference surface alone, by
the replica alone, and by their union, on the same pairs and trajectories.

\section{Results and discussion}\label{sec:results}

The results follow the questions a user of the workflow asks. Synthetic
channels with known ground truth first fix the meaning of the numbers, before
any measured surface is graded. The insert-to-part pairs then test whether
\RQd{} grades real replicas where the height correlation does not, and trace
what it responds to. The moulding run asks whether it follows the
degradation of a mould insert. The dimensions, the roughness and the print
texture read from the same scans give the complementary diagnosis that places
a deviation along the CAD-to-insert-to-part chain. Finally, the metric is
tested against established similarity measures and against its own analysis
settings, and its scope is stated.

\subsection{Synthetic characterisation of \texorpdfstring{\RQd}{RQd}}\label{sec:res-synthetic}

Before the experimental chain, the behaviour of \RQd{} was characterised on
the synthetic surfaces, analysed with
the same production code and fixed \SI{5}{\micro\meter} settings used
throughout.
Two references bracket the metric: identical surfaces return exactly $1$,
and an independently textured surface with the channel geometry removed
returns $-0.001{\pm}0.009$ (mean $\pm$ standard deviation over $30$
texture seeds).
Figure~\ref{fig:synthetic}(a)--(d) shows the reference channel alongside
progressively smoothed and locally defective variants. On
the reference geometry the sidewalls occupy about $5\,\%$ of the analysed cells
yet carry about $97\,\%$ of the gradient energy within the channel-interior
mask. This anticipates the wall dominance found on the measured channels.

Under progressively stronger wall smoothing, \RQd{} falls
monotonically, from $1.000$ to $0.872$, $0.683$, $0.551$, and $0.384$ at
smoothing scales $\sigma = 40$, $80$, $120$, and \SI{200}{\micro\meter}
(Fig.~\ref{fig:synthetic}(e)). \RQ{} falls under the same degradation but
flattens. Its slope peaks at about $0.04$ per \SI{10}{\micro\meter} for
$\sigma$ between $40$ and \SI{50}{\micro\meter} and is below $0.01$ per
\SI{10}{\micro\meter} beyond \SI{160}{\micro\meter}. It thus still reads
$0.604$ at \SI{200}{\micro\meter} for a channel whose walls have visibly gone
(Fig.~\ref{fig:synthetic}(c)), whereas \RQd{} is still losing $0.014$ per
\SI{10}{\micro\meter} there. Beyond $\sigma \approx \SI{120}{\micro\meter}$,
\RQ{} moves by $0.07$ in total and \RQd{} by $0.17$, consistent with the
saturation that the printed channels show. Because the curve in
Fig.~\ref{fig:synthetic}(e) is monotonic, a value measured on a printed
channel can in principle be read back as an equivalent wall-smoothing scale.

Smoothing confined to the wall corners and smoothing applied to the whole
field give nearly the same \RQd{} at the same $\sigma$
(Fig.~\ref{fig:synthetic}(f)). They agree within $0.015$ up to
\SI{100}{\micro\meter} and within $0.07$ at \SI{200}{\micro\meter}. In these
tests the metric thus responds to the loss of wall steepness whether it is
confined to the walls or spread over the field. The two modes produce the
same geometry at equal $\sigma$ and differ only in the microtexture, which
wall smoothing leaves intact and global blur removes. Below
\SI{100}{\micro\meter} the global blur scores marginally lower, because
blurring the texture slightly decorrelates it from the reference. Above
\SI{100}{\micro\meter} it scores higher. Once the walls are weak, the intact
texture of the wall-smoothed surface competes with them in the gradient map,
whereas the blurred surface keeps a clean, if faint, wall band. A heavily
degraded but still rough surface therefore scores lower than an equally
degraded smooth one, consistent with the noise result below.

The metric is exactly invariant to positive amplitude scaling, since Pearson
correlation is blind to a positive factor. For the same reason it stays high
under floor filling. With the opening width
held fixed, \RQd{} remains above $0.98$ up to $70\,\%$ depth loss, reads
$0.93$ at $80\,\%$ and $0.79$ at $85\,\%$, and falls to $0.43$ at
$90\,\%$ (Fig.~\ref{fig:synthetic}(g)). This last drop comes from the
channel-interior mask rather than from the walls. The filled channel is then
barely deeper than the mask level, so the mask keeps only about $6\,\%$ of
its cells. \RQ{}, also amplitude-invariant, is almost equally insensitive to a uniform loss of depth ($0.85$ at $80\,\%$, $0.50$ at $90\,\%$). Neither correlation is
therefore a depth check. \RQd{} grades the spatial correspondence of steep
and flat regions, not absolute height or depth, for which the extracted
dimensions and the $E_Z$ map remain necessary.

Added Gaussian noise with a standard deviation of \SI{20}{\micro\meter} costs
\RQd{} $0.04$, \SI{40}{\micro\meter} costs $0.18$ and \SI{80}{\micro\meter}
costs $0.53$. All these levels are far above the instrument's
vertical-resolution setting of \SI{0.5}{\micro\meter}. The dispersion of
repeated scans on the sidewalls was not measured and is absorbed in the
self-comparison of the rescanned inserts. \RQ{} is less noise-sensitive
($0.89$ at \SI{80}{\micro\meter}) because differentiation amplifies noise.
The largest isolated synthetic defect changes \RQd{} by less than $0.01$
while remaining conspicuous in the $E_d$ map, which is why the maps accompany
the scalar. Imposed rigid transformations strongly reduce the unregistered
correlation. The production registration, however, reduces this bias to at
most $0.002$; the seed-to-seed standard deviation is
at most $0.010$. All synthetic values are $\RQd(\SI{5}{\micro\meter})$ on the
generator's \SI{5}{\micro\meter} grid; the cell size is the strongest
numerical setting. The same fixed settings are used for every measured
surface in the sections that follow, so the synthetic and the experimental
values are directly comparable. Section~S2 of the supplementary material
gives the full characterisation.

\subsection{\texorpdfstring{\RQ{} and \RQd{}}{RQ and RQd} on the insert-to-part pairs}\label{sec:res-saturation}\label{sec:res-resin}

\begin{figure}[pos=!tb]
\centering
\includegraphics[width=\columnwidth]{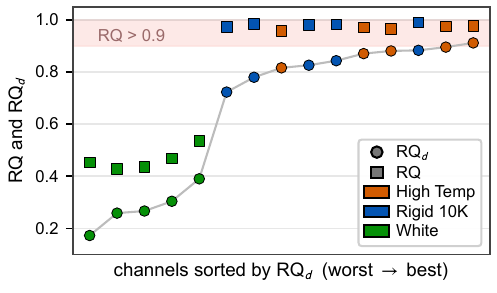}
\caption{$\mathrm{RQ}$ and $\mathrm{RQ}_{d}$ for the fifteen insert-to-part
comparisons, ordered by $\mathrm{RQ}_{d}$ and coloured by resin; the shaded
band marks $\mathrm{RQ} >$ 0.9.}
\label{fig:saturation}
\end{figure}

Across all fifteen insert-to-part comparisons, \RQd{} resolves replication
quality over almost its entire range. It runs from $0.17$ for the
worst-formed channels to $0.91$ for the best (Fig.~\ref{fig:saturation}; every pair in Table~S3.1). The height correlation \RQ{} does the
opposite. Once a channel has formed at all, \RQ{} collapses into a narrow
band against unity: the ten channels with $\RQ > 0.9$ span only $0.03$ in
\RQ{} yet $0.19$ in \RQd{}. The two metrics rank the channels only loosely
alike (rank correlation $0.72$). The height correlation can even reverse an ordering that the extracted cross-section geometry independently supports. The rim of the R10K \SI{600}{\micro\meter} part is
rounded by \SI{35}{\micro\meter}, against \SI{4}{\micro\meter} on its insert,
and its opening widens by \SI{57}{\micro\meter}, although its wall angle
differs by only $1.5^{\circ}$ (Table~S4.1). It scores $\RQ = 0.97$ but
$\RQd = 0.72$. The HT \SI{100}{\micro\meter} channel, reproduced to within
$0.6^{\circ}$ in wall angle, \SI{5}{\micro\meter} in width and
\SI{10}{\micro\meter} in corner rounding, scores $\RQ = 0.96$ and
$\RQd = 0.82$. On these data the height correlation confirmed that a channel
had formed but did not resolve how faithfully a well-formed channel was
reproduced. This is why \RQd{} is taken as the primary metric. The
insensitivity to shape, which differentiation removes, also distinguishes
\RQd{} from earlier correlation metrics, which correlate heights rather
than slopes.

The three resins span the quality range that the metric is meant to
discriminate (Table~\ref{tab:rq_summary}). The first-shot \RQd{} averages
$0.87$, $0.81$, and $0.28$ for HT, R10K, and White, respectively. For the
same comparisons the height correlation reads $0.97$, $0.98$, and $0.46$,
saturating near unity for the two resins that moulded well. \RQd{} separates HT from R10K, which the height correlation places together; both metrics separate them from the failed White.

\begin{table}[!htbp]
\centering
\caption{First-shot fidelity by resin, given as mean $\pm$ one standard
deviation over the five channel widths.}
\label{tab:rq_summary}
\setlength{\tabcolsep}{2pt}
\begin{tabularx}{\linewidth}{@{}l*{4}{>{\centering\arraybackslash}X}@{}}
\toprule
 & CAD$\to$Ins. & \multicolumn{2}{c}{Insert$\to$Part} & CAD$\to$Part \\
\cmidrule(lr){3-4}
Resin & $\mathrm{RQ}$ & $\mathrm{RQ}$ & $\mathrm{RQ}_{d}$ & $\mathrm{RQ}$ \\
\midrule
HT & 0.80\,$\pm$\,0.06 & 0.97\,$\pm$\,0.01 & 0.87\,$\pm$\,0.04 & 0.81\,$\pm$\,0.05 \\
R10K & 0.82\,$\pm$\,0.10 & 0.98\,$\pm$\,0.01 & 0.81\,$\pm$\,0.06 & 0.83\,$\pm$\,0.10 \\
White     & 0.68\,$\pm$\,0.18 & 0.46\,$\pm$\,0.04 & 0.28\,$\pm$\,0.08 & 0.61\,$\pm$\,0.10 \\
\bottomrule
\end{tabularx}
\vspace{-8pt}
\end{table}

\begin{figure*}[!tb]
\centering
\includegraphics[width=\textwidth]{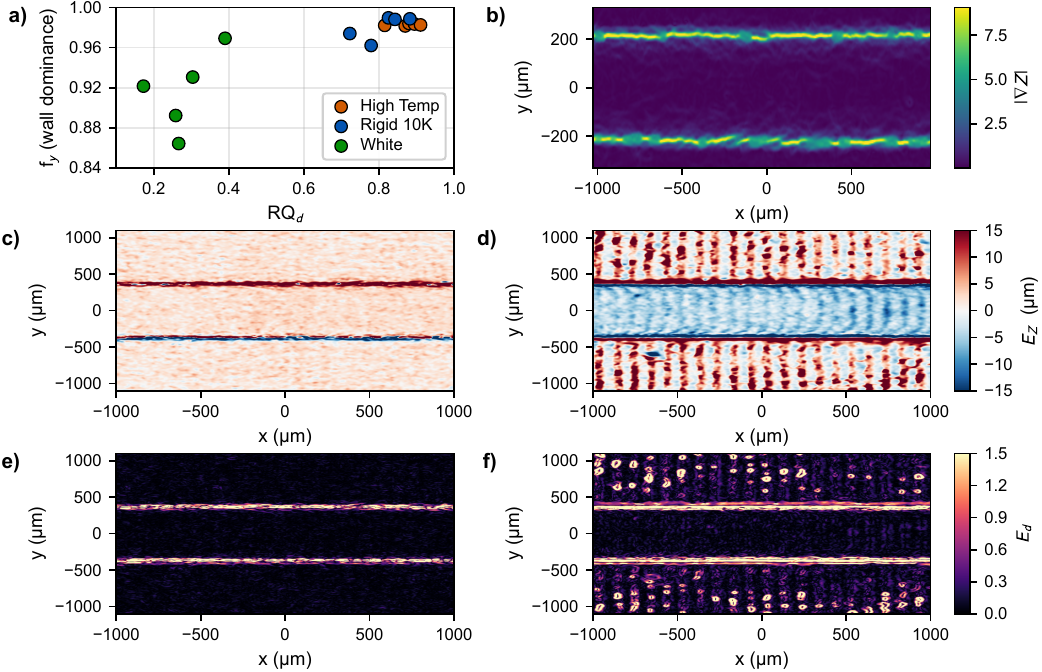}
\caption{Wall dominance of the gradient magnitude and per-pixel error maps.
(a)~Wall-dominance ratio $f_y$ against $\mathrm{RQ}_{d}$ for the fifteen
insert-to-part pairs, coloured by resin. (b)~Gradient magnitude $|\nabla Z|$
over a central segment of a High Temp \SI{300}{\micro\meter} channel.
(c,d)~Signed height error $E_Z$ and (e,f)~gradient-magnitude error $E_d$ of a
High Temp (c,e) and a Rigid 10K (d,f) \SI{600}{\micro\meter} channel at the
first shot.}
\label{fig:mech_maps}
\vspace{-1pt}
\end{figure*}

Even without replication loss, \RQd{} for real surfaces stays below unity.
Comparing each insert with itself, scanned before and after a single
shot, returns $\RQd = 0.88 \pm 0.02$ for HT and $0.90 \pm 0.02$ for R10K
over the five widths (mean $\pm$ standard deviation). This is an
intermediate precision condition rather than a repeatability condition
\citep{vim2012}: the same procedure and instrument, but with demounting,
one moulding shot and remounting between the scans. The comparison therefore
contains any change of the insert in that shot as well as the measurement.
For HT the wall geometry records such a change. In the first shot its walls
straightened and tilted: the wall angle rose from $7.2^{\circ}$ to
$12.7^{\circ}$, the rounding of the floor and rim corners fell by $2$ to
\SI{14}{\micro\meter} and the opening widened by $5$ to
\SI{12}{\micro\meter}, while the floor width and the depth stayed within
\SI{5}{\micro\meter} (Table~S6.3). The part moulded in that shot still
carries the as-printed shape. After the first shot the HT insert held
its shape, and its scans after $1$, $5$, $10$ and $20$ shots compare with one
another at $0.92$ to $0.95$ (Table~S6.4). These later self-comparisons are
effectively correlations between independent measurements of the same
surface. They are set by the decorrelation of fine microtexture between scans
and by residual registration error rather than by any replication loss. Both
contributions are expected: height fluctuations in repeated optical
measurements grow with the local gradient \citep{lemesle2023}, so the
sidewalls that carry the \RQd{} signal are also where scan-to-scan noise is
largest. In the measurement-study model of Liu et al.~\citep{liu2024}, an
observed similarity is the true similarity plus a measurement term, and an
insert compared with its own unchanged state estimates that term directly.
First-shot HT replication ($0.87$) therefore lies below the rescan level
observed under the present protocol ($0.92$ to $0.95$): even the
best-replicating resin loses some sidewall fidelity in moulding. On the
synthetic surfaces the analysis itself contributes little: the registration
bias is at most $0.002$ and the seed-to-seed standard deviation at most
$0.010$. The HT insert, unchanged after the first shot, nevertheless moves by
up to $0.02$ between shots at a single width (standard deviation $0.013$ over
shots and widths). Differences in \RQd{} smaller than about $0.02$ are
therefore not interpreted. This bound rests on the analysis and on the
same-tool scatter of one insert; it is not a measurement uncertainty.

\subsection{Physical interpretation}\label{sec:res-sidewall}\label{sec:res-errormaps}

The discriminating power of \RQd{} comes from the sidewalls. The gradient
correlation splits into two directional components, \RQdy{} across the
channel and \RQdx{} along it, and \RQd{} closely follows \RQdy{}: the two
agree to a correlation of $0.97$ (mean absolute difference $0.07$).
Geometrically, the across-channel gradient is concentrated at the walls
(Fig.~\ref{fig:mech_maps}(a,b)). The wall-dominance ratio \fy{} exceeds
$0.85$ for every channel. It reaches about $0.99$ for the well-replicated
ones and stays above $0.86$ even for the poorly replicated White channels.
For the channels studied, then, \RQd{} behaves as a measure of sidewall
fidelity, for all three resins. Sidewall form is also what a single width or
depth ratio does not capture, and it is where independent studies locate
replication loss. Loss concentrates at the steepest features of the master,
in the moulding of rough surfaces \citep{perris2023} as in embossed
microchannels \citep{yildirim2024}. Across a set of eleven areal parameters,
the slope statistics $S_{dq}$ and $S_{dr}$ are the ones that transfer worst
from a steel tool to the moulded part \citep{tosello2026}. The sidewall thus
appears to be both where replication loss tends to concentrate and where the
metric, by its construction, is most sensitive. This distinguishes it from a
general similarity index. It is also why the exact definition of the
channel-interior region matters little for well-formed channels: the walls
dominate the gradient energy under every mask tested.

Where the walls depart is shown by the error maps. \RQd{} is deliberately
a global grade of distributed fidelity; the synthetic study showed that a
severe but localised defect barely moves it. The per-pixel maps computed by
the software supply the complementary local view
(Fig.~\ref{fig:mech_maps}(c)--(f)). For the best-replicated channel (HT,
\SI{600}{\micro\meter}, $\RQd = 0.91$; panels (c) and (e)), the signed height
error $E_Z$ is small-amplitude and spatially random. The gradient-magnitude
error $E_d$ is confined to two thin sidewall lines. For the lowest-fidelity
opaque channel (R10K, \SI{600}{\micro\meter}, $\RQd = 0.72$; panels (d) and
(f)), the sidewall bands thicken and turn asymmetric. The floor departs
systematically, and the surrounding field carries a periodic mismatch
inherited from the print texture. The scalar provides the global grade and
the maps localise the deviations; together they capture both distributed
wall degradation and localised damage. The maps and \RQd{} derive from the
same registered topographies, so the maps localise the grade rather than
validate it independently. The check that does not share this construction
is the extracted cross-section geometry of Table~S4.1. The translucent White
parts cannot be mapped because their steep sidewalls drop out optically;
their penalty is carried by the channel-interior aggregate of
Table~\ref{tab:rq_summary}.

\subsection{Tracking insert degradation}\label{sec:res-lifetime}

\begin{figure*}[!tb]
\centering
\includegraphics[width=\textwidth]{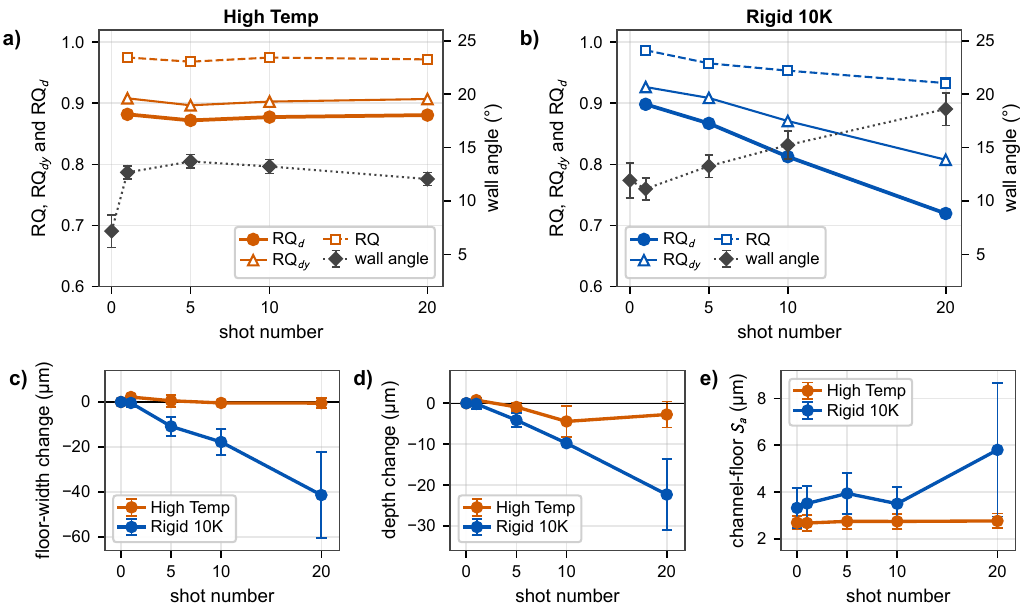}
\caption{Insert degradation over the moulding run. (a,b)~$\mathrm{RQ}_{d}$,
its across-channel component $\mathrm{RQ}_{dy}$ and $\mathrm{RQ}$ between each
insert at shot 0 and at shot $N$ (left axis), with the wall angle (right
axis), for (a)~High Temp and (b)~Rigid 10K. (c)~Floor-width and
(d)~depth change from shot 0, taken per width. (e)~Channel-floor $S_a$.
Values are means over the five widths; error bars are standard deviations.\looseness=-1}
\label{fig:lifetime}
\vspace{-1pt}
\end{figure*}

A replication metric for rapid tooling should also track how long the tool
maintains its quality. Following a measure across a run is established
practice. Channel dimensions have been followed to 100 and 500 shots
\citep{zhang2017, convery2021, menezes2024}, and a profile correlation ratio
to 30 \citep{perris2023}. In several of these records the measure held
constant to the last part \citep{zhang2017, convery2021, perris2023}. The
question is whether the measure responds to the degradation that does occur,
which Stampone et al.~\citep{stampone2024} saw by eye as a gradual change
of channel shape at constant channel depth. Each insert is graded against
its own as-printed state; grading against the moulded part instead would
hide the drift, because each part faithfully copies the current state of
its insert.
Figure~\ref{fig:lifetime}(a,b) shows the two resins that survived the run
(per-width values in Tables~S6.1 and S6.2). HT settled at the first shot and
then held. Against its as-printed state it stays within $0.01$ of $0.88$, the
value it reached at the first shot, through shot $20$, and its later states
compare with one another at $0.92$ to $0.95$. R10K drifts monotonically from $0.90$ to $0.72$; drift here denotes
the progressive change of the tool itself, not of the instrument. The height
correlation is insensitive to the drift that matters. It stays near $0.97$
for HT and falls only from $0.99$ to $0.93$ for R10K, a change that would
not by itself signal a degrading tool. White is not shown: it reached end of
life within the first five shots, its slender channel-forming ridges
shearing off. Only that catastrophic feature loss registers in the height
correlation ($0.78$ to $0.65$), as it does in \RQd{} ($0.79$ to $0.75$
before the record stops). This bears on the flat height-based records in the
literature. A profile correlation ratio of $0.74$, $0.75$ and $0.74$ at
shots 10, 20 and 30 \citep{perris2023} is consistent with a stable tool or
with an insensitive measure. A shape-sensitive metric is needed to tell a
stable tool from a degrading one.

\begin{figure*}[!tb]
\centering
\includegraphics[width=\textwidth]{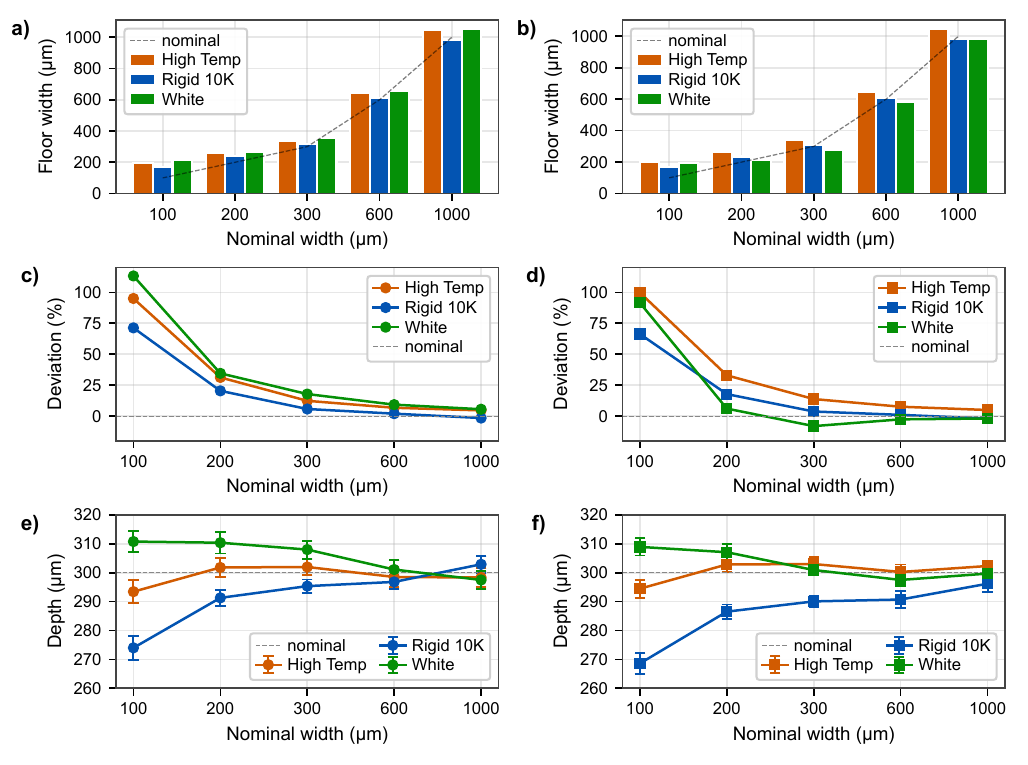}
\caption{Channel dimensions of the as-printed inserts (a,c,e) and their first
moulded parts (b,d,f), by resin: (a,b)~floor width, (c,d)~its deviation from
nominal and (e,f)~channel depth. Nominal values are dashed; error bars in
(e,f) are the along-channel standard deviation of the per-slice depths.}
\label{fig:oversizing}
\vspace{-6pt}
\end{figure*}

The drift has the geometric form the sidewall-sensitivity argument predicts,
and the panels make the chain of evidence explicit. In R10K
(Fig.~\ref{fig:lifetime}(b)) the across-channel component \RQdy{} falls in
step with \RQd{}, from $0.93$ to $0.81$, while the mean wall angle grows from
$11.9^{\circ}$ as printed to $18.6^{\circ}$ at shot $20$. In HT
(Fig.~\ref{fig:lifetime}(a)) \RQdy{} holds near $0.90$ and the wall angle,
after its step in the first shot, stays between $12.1^{\circ}$ and
$13.7^{\circ}$. The R10K walls wear to a shallower slope as a whole, their
floor-corner rounding even falling from $25$ to \SI{17}{\micro\meter}, while
the height correlation holds above $0.93$. Here, tool wear appeared as a
sidewall phenomenon that the slope-sensitive metrics registered and the
height correlation did not. The drift is not uniform across the channel
sizes. The \SI{100}{} to \SI{600}{\micro\meter} R10K channels fall to between
$0.64$ and $0.73$ by shot $20$. The \SI{1000}{\micro\meter} channel, the widest and
shallowest, is the most durable at $0.82$. HT is flat at every width, all
five staying within $0.02$ of their shot-1 values.

The dimensional and roughness measures then quantify the consequence
(Fig.~\ref{fig:lifetime}(c)--(e)). By shot $20$ the R10K floors have narrowed
by $41 \pm 19$~µm (mean $\pm$ standard deviation over the widths; $16$ to
\SI{63}{\micro\meter}, the \SI{200}{} and \SI{300}{\micro\meter} channels
losing most). Their depth has fallen by $22 \pm 9$~µm. The HT floor widths, by
contrast, stay within \SI{5}{\micro\meter} and the HT depths within
\SI{10}{\micro\meter} of their as-printed values throughout the run. The channel-floor roughness
follows the same split. HT holds $S_a$ near \SI{2.7}{\micro\meter}
throughout, while the R10K mean rises from $3.3$ to \SI{5.8}{\micro\meter}.
The rise is carried by the three narrowest channels ($7$ to
\SI{9}{\micro\meter} at shot $20$); the \SI{600}{} and
\SI{1000}{\micro\meter} floors do not roughen. The peak-to-valley height
$S_z$ of those narrow R10K floors grows from about $27$ to between $57$ and
\SI{71}{\micro\meter}. HT stays near \SI{21}{\micro\meter}, except for a
single localised value of \SI{53}{\micro\meter} at the
\SI{600}{\micro\meter} channel at shot $20$. This value is consistent with
local damage preceding the eventual fracture of the insert body, but it is
not by itself an early-warning criterion.

Read together, the panels track the degradation at several levels of
detail. \RQd{} is the first indicator to move. It registers the R10K drift
at shot $5$ ($0.90$ to $0.87$, beyond the $0.02$ interpretation bound), when the floor has narrowed by \SI{11}{\micro\meter}
and the roughness shows no sustained rise. It also grades how far the tool has
departed from its as-printed state. \RQdy{} and the wall angle locate the
change at the sidewalls, and the per-width values show which features go
first. The width, depth and roughness measures give the dimensional and
surface consequences that a specification can be written against. No single one would have sufficed: the height correlation stays above
$0.93$ throughout, and a sustained roughness rise appears only at shot $20$.

\subsection{Dimensional fidelity}\label{sec:res-print}

Because \RQd{} discards absolute size, the dimensions extracted from the
same scans are read alongside it as the complementary diagnosis. The chain
resolves into two steps, CAD to insert (printing) and insert to part
(moulding). The dominant dimensional deviation is introduced at the print
stage, not at moulding (Table~S4.1 in the supplementary material). Channel
width is systematically oversized in printing, and the effect is strongest
for the narrowest channels. The nominal \SI{100}{\micro\meter} channel prints
to about $195$, $213$, and \SI{171}{\micro\meter} in HT, White, and R10K,
respectively. Moulding, by contrast, is dimensionally faithful for the resins
that mould well. The HT and R10K parts track their own inserts' floor width
to within \SI{6}{\micro\meter} (at most $2.9\,\%$), with one sign per resin:
the HT parts come out $3$ to \SI{5}{\micro\meter} wider than their inserts
and the R10K parts $4$ to \SI{6}{\micro\meter} narrower, at every width.
Channel depth stays close to the \SI{300}{\micro\meter} nominal ($269$ to
\SI{311}{\micro\meter} across all inserts and parts). The White parts, whose
replication fails, depart from their inserts in floor width by up to
\SI{78}{\micro\meter} (Table~S4.1). Two
practical implications are suggested (Fig.~\ref{fig:oversizing}). First,
small-feature fidelity points to compensating the width at the print stage,
where the deviation originates, rather than by tuning the moulding cycle.
Such compensation is an established practice in precision additive
manufacturing \citep{afazov2017} and has been applied to SLA mould inserts as
fixed height and width offsets \citep{convery2021}. Second, since moulding
copied whatever the print produced for these resins, optimisation effort
appears best directed at the print step. Resolved by resin
(Fig.~\ref{fig:oversizing}(a)--(d)), the oversizing is systematic rather than
material-specific. It decays with nominal width in the same way for all
three resins, from $71$ to $113\,\%$ at \SI{100}{\micro\meter} to within
$\pm 6\,\%$ at \SI{1000}{\micro\meter}. White prints the widest and the
glass-filled R10K the narrowest; R10K comes within $2\,\%$ of nominal at
\SI{600}{\micro\meter} and prints $1.5\,\%$ narrow at \SI{1000}{\micro\meter}.
Each moulded part follows its own insert rather than the nominal.

Depth behaves differently from width (Fig.~\ref{fig:oversizing}(e,f)). It
varies much less with nominal width, because the feature height is set by
the build direction rather than by the in-plane resolution that governs the
printed width. It carries no systematic oversizing. The printed inserts span
$274$ to \SI{311}{\micro\meter} about the \SI{300}{\micro\meter} nominal,
with resin means of $299$ (HT), $306$ (White), and \SI{292}{\micro\meter}
(R10K). The clearest departure is the R10K \SI{100}{\micro\meter} channel,
whose insert and part read $274$ and \SI{269}{\micro\meter}. The narrowest,
highest-aspect feature of the glass-filled resin is the one that fills least
completely. Moulding again copies whatever the print produced: the part
depth tracks its own insert to within \SI{7}{\micro\meter} for every
condition. Depth thus supports the conclusion drawn from width, that the
print stage sets the dimensional envelope. The longitudinal variability of
depth is small, at most \SI{4}{\micro\meter} across the conditions that
satisfy the depth quality check. The single excluded condition is a White
part whose sidewall dropout leaves too few valid slices for a meaningful
spread. Because depth and width respond to different aspects of the process,
neither alone characterises the feature. This is the practical case for a
metric that grades the whole cross-section at once, from the same scan that
yields the dimensions.

\subsection{Areal roughness transfer}\label{sec:res-roughness}

The three resins present distinct as-printed channel-floor finishes
(Table~\ref{tab:roughness}; per channel in Table~S5.1). Among the
inserts, HT prints the smoothest floor ($S_a = 2.69$, range
$2.29$--\SI{3.08}{\micro\meter} over the five widths). White and R10K are
similar but rougher ($3.22$ and \SI{3.32}{\micro\meter}). R10K also shows
the largest width-to-width variation (range $2.62$--\SI{4.59}{\micro\meter}).
Its highly glass-filled formulation may contribute, since dispersed filler
particles can modify resin rheology, light scattering, and local surface
formation during printing. This interpretation remains tentative, because
filler content was not varied independently. The flat field follows the same
resin ordering as the floor ($S_a$ of $2.33$, $4.38$, and
\SI{3.35}{\micro\meter} for HT, R10K, and White, each with a standard
deviation below \SI{0.3}{\micro\meter} over the five widths). The smooth HT
surface is consistent with its vanishing layer-line signature.

\begin{table}[!htbp]
\centering
\caption{Areal roughness $S_a$ (\si{\micro\meter}) of the channel floor of the
as-printed inserts and their first moulded parts, and of the insert field;
mean $\pm$ one standard deviation over the five channel widths.}
\label{tab:roughness}
\begin{tabularx}{0.9\linewidth}{@{}l*{3}{>{\centering\arraybackslash}X}@{}}
\toprule
 & \multicolumn{2}{c}{Floor} & Field \\
\cmidrule(lr){2-3}
Resin & Insert & Part & Insert \\
\midrule
HT & 2.69\,$\pm$\,0.29 & 2.35\,$\pm$\,0.34 & 2.33\,$\pm$\,0.15 \\
R10K & 3.32\,$\pm$\,0.86 & 3.36\,$\pm$\,0.67 & 4.38\,$\pm$\,0.27 \\
White     & 3.22\,$\pm$\,0.40 & 3.31\,$\pm$\,0.38 & 3.35\,$\pm$\,0.19 \\
\bottomrule
\end{tabularx}
\end{table}

Moulding transfers this texture nearly one to one
(Table~\ref{tab:roughness}). The part channel-floor $S_a$ tracks its insert
for every resin: $2.35$ against \SI{2.69}{\micro\meter} (HT), $3.36$
against \SI{3.32}{\micro\meter} (R10K), and $3.31$ against
\SI{3.22}{\micro\meter} (White). The per-channel transfer ratio is $0.98$ on
average ($0.87$, $1.02$, and $1.03$ by resin, the part marginally smoother
only for HT). The \mbox{root-mean-square} height $S_q$ transfers in the same
proportion ($0.88$, $1.00$ and $1.09$ by resin). The maximum height $S_z$
does so for HT and R10K ($0.96$ and $0.83$) but doubles for the failed White
parts. Surface finish therefore appears to be inherited from the printed
insert rather than generated in moulding, consistent with height-based
texture-transfer studies \citep{piccolo2021, vera2018}. \RQd{} adds the
shape information that roughness statistics do not provide. A floor-finish
specification is therefore best placed on the insert.

\subsection{Layer-line periodicity and build orientation}\label{sec:res-layerlines}

\begin{figure}[pos=!tb]
\centering
\includegraphics[width=\columnwidth]{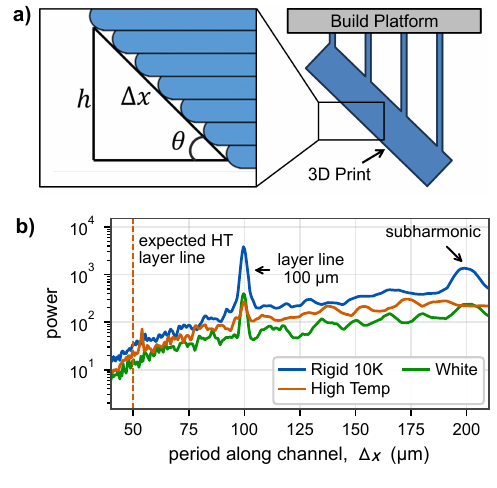}
\caption{Layer lines and build orientation. (a)~Layer thickness $h$, build
angle $\theta$ and period $\Delta x$. (b)~Along-channel power spectra of the
printed inserts, normalised by the median of each full spectrum.}
\label{fig:layerlines}
\vspace{-5pt}
\end{figure}

The SLA layering leaves a deterministic texture that encodes the build, and
spectral analysis of areal topography \citep{jacobs2017} recovers it. The
along-channel spectrum of the insert floor and field shows a sharp peak at
$\Delta x = \SI{100}{\micro\meter}$ for the \SI{50}{\micro\meter}-layer
resins, $16$ and $7$ times above the local background (the running median
over \SI{25}{\micro\meter} of period) for R10K and White
(Fig.~\ref{fig:layerlines}(b)). A layer of thickness $h$ built at an angle
$\theta$ to the platform intersects the surface every
$\Delta x = h/\sin\theta$ along the channel (Fig.~\ref{fig:layerlines}(a)),
so the peak returns the build orientation as $\theta = \arcsin(h/\Delta x)$:
$30.1^{\circ}$ for both resins against the $30^{\circ}$ set in the slicer.
A weaker subharmonic at $2\Delta x = \SI{200}{\micro\meter}$ is consistent
with a two-layer period: the hatch direction alternates between layers, so
the surface repeats every second layer as well. The \SI{25}{\micro\meter}-layer
HT print shows no dominant line. Its strongest features, at $27$, $36$, $54$
and \SI{100}{\micro\meter}, stand only two to three times above the
background. The one at \SI{54}{\micro\meter} lies where a
\SI{25}{\micro\meter} layer at $27.6^{\circ}$ would place it but cannot be
singled out: the smoother print reduces the layer-line signature to the
level of the background, and the orientation is not recoverable. The
layer-line spectrum thus serves as a print-quality fingerprint that records
the build orientation and gauges print smoothness, from the same scan used
for \RQd{} and at no extra measurement cost. It complements studies that
relate the stair-step texture of printed tooling to the build orientation
through roughness \citep{keane2025} or cusp-height geometry
\citep{poskus2023}. The peak ratios are tabulated in Table~S9.1.

\subsection{Benchmark against other similarity measures}\label{sec:res-benchmark}

A natural question is whether an existing similarity measure would have
graded these data equally well. Table~\ref{tab:benchmark} (per pair in
Table~S7.1) answers it with the height correlation \RQ{} and the reference
measures $1 - D_s$, SSIM, GMSD and GMSM. The test sets are the fifteen
insert-to-part pairs, the synthetic wall-smoothing series and the
tool-degradation trajectories. Over the fifteen pairs every measure ranks the
channels broadly as \RQd{} does (Spearman $|\rho_s|$ from $0.72$ to $0.90$),
because the failed White channels dominate any ranking. The measures
separate on the ten well-formed channels. \RQd{} spreads them over $0.19$,
whereas the height correlation ($0.03$), SSIM ($0.04$), $1 - D_s$ ($0.07$)
and the two gradient-similarity poolings ($0.05$ and $0.02$) compress them
into a few hundredths. Every measure except \RQd{} and $1 - D_s$ places the
R10K \SI{600}{\micro\meter} channel above the HT \SI{100}{\micro\meter} channel,
although its rim has rounded by about \SI{30}{\micro\meter} relative to its
insert.
The extracted sidewall geometry (Table~S4.1) and the error maps of
Fig.~\ref{fig:mech_maps} contradict this reversal. On the synthetic series
all respond monotonically to wall smoothing except $D_s$. \RQd{}, however,
falls to $0.38$ at the largest scale, where SSIM stays at $0.80$ and GMSM at
$0.92$. Over the moulding run, the same-tool spread $s_{\mathrm{HT}}$ is the
standard deviation of each measure on the HT insert, which held its shape
after the first shot, across shots 1 to 20 and the five widths. The
drift of the R10K insert is resolved at fourteen times $s_{\mathrm{HT}}$ by
\RQd{} and at fifteen times by $D_s$. The height correlation and SSIM
resolve it at six to seven times, and GMSD and GMSM at two to three.\looseness=-1

\begin{table*}[!tb]
\centering
\caption{Benchmark of similarity measures against \RQd{}.}
\label{tab:benchmark}
\setlength{\tabcolsep}{3.5pt}
\begin{tabular}{@{}llccccccc@{}}
\toprule
 & & \multicolumn{4}{c}{Fifteen insert-to-part pairs} & Synthetic & \multicolumn{2}{c}{R10K drift, shots 1 to 20} \\
\cmidrule(lr){3-6} \cmidrule(lr){7-7} \cmidrule(lr){8-9}
Measure & Bounded & Range & Spread & $\rho_s$ & Order & $\sigma = \SI{200}{\micro\meter}$ & Change & Change$/s_{\mathrm{HT}}$ \\
\midrule
$\mathrm{RQ}$, height correlation & yes & 0.43 to 0.99 & 0.03 & 0.72 & no & 0.60 & \num{-0.05} & 7 \\
\RQd{}, this work & yes & 0.17 to 0.91 & 0.19 & 1 & yes & 0.38 & \num{-0.18} & 14 \\
$1 - D_s$, NIST difference & no & \num{-43} to 0.96 & 0.07 & 0.90 & yes & 0.30 & \num{-0.26} & 15 \\
SSIM, structural similarity & yes & 0.75 to 0.99 & 0.04 & 0.87 & no & 0.80 & \num{-0.08} & 6 \\
GMSD, gradient-similarity deviation & yes & 0.04 to 0.28 & 0.05 & \num{-0.88} & no & 0.18 & \num[retain-explicit-plus]{+0.02} & 2 \\
GMSM, gradient-similarity mean & yes & 0.88 to 0.99 & 0.02 & 0.82 & no & 0.92 & \num{-0.02} & 3 \\
\bottomrule
\end{tabular}
\vspace{-1pt}
\end{table*}

Two conclusions follow. First, the two image-quality indices behave as
their design objectives predict. Both rate the perceived quality of an image
against a pristine reference. SSIM compares local means, variances and
covariance, so on height maps it inherits the saturation of the height
correlation. GMSD pools the pointwise gradient similarity by its standard
deviation because, for perceived quality, the spatial non-uniformity of a
distortion matters more than its mean. A tool that degrades uniformly along
its walls, as R10K does, therefore changes it little, which is the case a
wear metric should register. Neither behaviour is a defect of the index; it
follows from an objective different from replication grading. Second, $D_s$
is the strongest alternative for the moulding run, but for a different
reason. It is sensitive to amplitude by construction \citep{vorburger2015},
and the R10K drift is accompanied by a loss of depth and width. It is
unbounded (the White pairs reach $D_s = 44$) and does not separate the
well-formed channels. Its amplitude sensitivity would also register a
uniform shrinkage of the part, which the correlations by construction
ignore. \RQd{} is the only measure in the set that is bounded, grades among
good channels, agrees with the sidewall geometry on the example pair and
resolves the drift.

\begin{figure*}[!tb]
\centering
\includegraphics[width=\textwidth]{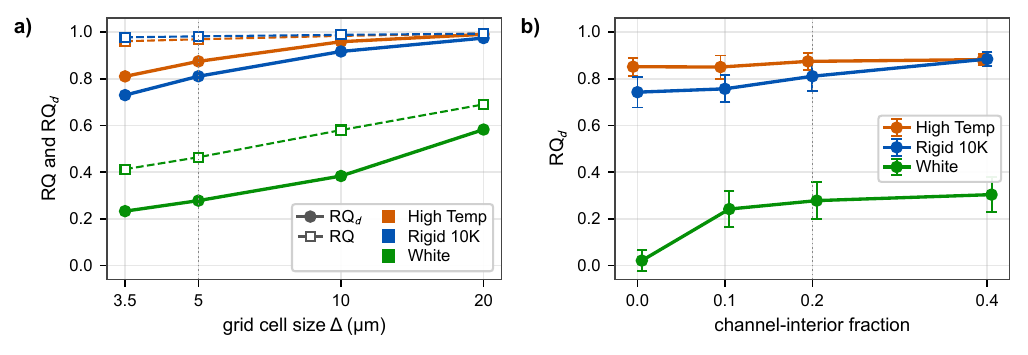}
\caption{Sensitivity to the analysis settings on the fifteen insert-to-part
pairs, each setting varied alone; dotted lines mark the defaults.
(a)~\RQd{} (filled) and $\mathrm{RQ}$ (hollow) against the grid cell size
$\Delta$, mean per resin over the five widths.
(b)~\RQd{} against the channel-interior fraction, mean $\pm$ standard
deviation over the five widths.}
\label{fig:sensitivity}
\end{figure*}

\RQd{} extends the one-axis slope residual of Loaldi
et al.~\citep{loaldi2021} to an isotropic, bounded correlation. It differs from the height-based correlation metrics of replication accuracy \citep{vorburger2011, kumar2019, perris2023} and topography matching \citep{zheng2020, wang2021sim, jiang2021} in the shape sensitivity that lets it grade quality where a height correlation saturates. Gradient-based
similarity is also established outside surface metrology. Image-quality
assessment uses SSIM \citep{wang2004ssim} and GMSD \citep{xue2014gmsd}.
Image registration uses the mean of the two directional gradient
correlations as a similarity measure \citep{penney1998}, the construction of
$\mathrm{RQ}_{dx}$ and $\mathrm{RQ}_{dy}$. \RQd{} is closest to these, but
differs in what is pooled as well as in what is compared. GMSD forms a
pointwise similarity of two gradient-magnitude maps and pools it by its
standard deviation. It therefore measures how uniformly an image is degraded
rather than by how much. \RQd{}, by contrast, is the Pearson correlation of
the two magnitude maps, invariant to a positive affine rescaling of either. It is
evaluated over a channel-interior mask on registered height fields and
interpreted against the geometry of a microchannel. The contribution claimed
here is the definition, characterisation, and interpretation of such a
correlation as a replication-quality grade, together with the tool that
computes it, rather than the gradient operator. The benchmark of
Table~\ref{tab:benchmark} isolates that combination. None of its ingredients
reproduces it alone, and each alternative falls short in at least one
respect. The benchmark also marks the one case that \RQd{} does not cover.
A uniform loss of depth at constant opening leaves it unchanged
(Fig.~\ref{fig:synthetic}(g)); there, the amplitude-sensitive $D_s$, or the
extracted depth itself, is the measure to read.

\subsection{Sensitivity to the analysis parameters}\label{sec:res-sensitivity}

Slope quantities depend on the sampling interval at which the slope is
taken \citep{lemesle2023}. The numerical values reported above are therefore
those of $\RQd(\Delta)$ at $\Delta = \SI{5}{\micro\meter}$ and the fixed
analysis settings. Of the settings varied (Fig.~\ref{fig:sensitivity}; all
sweeps in Tables~S8.1 and S8.2), the grid cell size is the dominant
parameter. Coarsening it from \SI{5}{} to
\SI{20}{\micro\meter} raises the mean \RQd{} of the fifteen pairs from
$0.65$ to $0.85$, and that of the HT channels from $0.87$ to $0.99$. The
fine texture that decorrelates between two scans is then averaged away.
Refining the cell to \SI{3.5}{\micro\meter}, the resolution limit, lowers
the mean to $0.59$. This point is evaluated on clouds thinned to
\SI{3}{\micro\meter} rather than \SI{5}{\micro\meter} voxels, which would
leave two fifths of the \SI{3.5}{\micro\meter} cells empty. The ranking is
stable throughout: Spearman $\rho_s \geq 0.94$ against the default, and the
HT \SI{100}{\micro\meter} channel stays above the R10K
\SI{600}{\micro\meter} channel at every cell size. The grading power,
however, grows monotonically as the cell shrinks. The spread of \RQd{} among
the ten well-formed channels is $0.23$ at \SI{3.5}{\micro\meter}, $0.19$ at
\SI{5}{\micro\meter} and $0.03$ at \SI{20}{\micro\meter}, where a coarse
gradient correlation saturates. Nothing is gained by coarsening the grid. At \SI{5}{\micro\meter}, the finest cell the instrument's resolution fills without interpolation, the grid sits at the fine end of the
usable range. The R10K drift is resolved at every cell size, at $9.9$ to
$24$ times the HT spread.\looseness=-1

The channel-interior fraction matters at its extremes. With no mask the flat
field dominates and the failed White channels fall to $0.02$. At $0.4$ the
walls are largely excluded and the R10K channels catch up with HT ($0.884$
against $0.882$). Between $0.1$ and $0.2$ the resin means move by less than
$0.06$. A Gaussian filter of \SI{5}{\micro\meter} changes nothing, because
it is the $\sigma = \Delta$ filter of the definition itself. A
\SI{10}{\micro\meter} filter lifts every value by about $0.09$ without
changing the order. Finally, a one-cell (\SI{5}{\micro\meter})
misregistration across the channel changes the mean of the fifteen pairs by
less than $0.005$ and the resin means by at most $0.02$. Two cells cost
$0.03$. Four cells (\SI{20}{\micro\meter}) decorrelate the ranking from the
default ($\rho_s = 0.51$) and reverse the example pair, the
\mbox{steep-walled} HT channels suffering most. Two conclusions follow.
\RQd{} is comparable only at equal cell size, so $\Delta$, fixed here at
\SI{5}{\micro\meter} by the instrument's lateral resolution, belongs to the definition. Registration to within about one grid cell is a precondition, met here by the pipeline. Both conditions are stated alongside every value reported in this paper.

Among the ten well-formed channels, the alternative channel-interior masks
(Table~S8.3) changed \RQd{} by at most
$0.048$ and preserved the ranking (Spearman $\rho_s \geq 0.93$ against the
default). The HT \SI{100}{\micro\meter} channel remained above the R10K
\SI{600}{\micro\meter} channel under every mask, and the R10K moulding-run
trajectory remained monotonic (Table~S8.4). Differences became substantial
only for the failed White channels. For these, reference-only regions
accounted for $8$ to $22\,\%$ of the union of the two channel masks.
Including these regions reduced \RQd{} from $0.17$ to $0.39$ under the
default mask to $-0.006$ to $0.096$ under the reference-defined mask. The
intersection mask therefore grades gradient correspondence within the shared
channel-interior footprint and under-weights geometry present on only one
surface. Such failures remain evident in the dimensions: the White parts depart from their inserts in floor width by up to
\SI{78}{\micro\meter}.

\subsection{Scope and limitations}\label{sec:res-scope}

Although demonstrated here on injection moulding, the metric is not tied to
it. It operates on any pair of registered profilometry point clouds. It can
therefore in principle be applied to soft lithography, embossing,
nanoimprint and \mbox{thermoforming}, wherever a master is copied into a
polymer and the fidelity of microscale shape is to be graded. The height
maps may come from any instrument that returns an areal topography; these
applications remain to be demonstrated. Its open-source,
\mbox{in-browser} implementation may lower the barrier to adoption for
laboratories without dedicated metrology software or a licence for it.

Several limitations bound the present study. \RQd{} requires the insert and
part scans to be registered to a common frame to within about one grid cell.
It is defined at a stated cell size, here $\Delta = \SI{5}{\micro\meter}$,
across which values are not comparable. The test artefact has straight channels only; curved, branched and free-form microstructures remain to be tested. The resin set is small
($n=3$), chosen to span a quality range rather than to sample the material
space. With one insert per condition and discrete measurement checkpoints,
the degradation observations are trends within the tested set rather than
population estimates. No repeated scans without an \mbox{intervening}
moulding cycle were acquired. Measurement \mbox{repeatability} is therefore
bounded only from above, by the intermediate-precision self-comparison of
the rescanned inserts. Replicate inserts and repeated scans would be needed
to separate genuine condition effects from fabrication and measurement
scatter. The $\pm$ values are standard deviations over widths or seeds, and
the longitudinal spreads are within-scan variation along a channel. The
agreements between insert, part and design dimensions are differences of
measured values. Measurement uncertainty was not evaluated, and the $0.02$
interpretation bound comes from the analysis and the same-tool scatter of
one insert. The CAD values are design
values without tolerance limits, and no conformity assessment is made.
Optical profilometry drops out on the steepest sidewalls of the translucent
White parts: the valid fraction averages $92\,\%$ over the comparisons but
falls to $66\,\%$ there. Counting only the cells measured on both surfaces
changes \RQd{} by at most $0.016$. The per-pixel error maps therefore use the opaque
resins, and White is graded through the channel-interior aggregate; an
opaque part material would remove this restriction.\looseness=-1

\section{Conclusions}\label{sec:conclusions}

Replication quality in mould-based microfabrication is usually judged from a
width or depth ratio at one cross-section. Such a ratio misses the sidewall
shape and depends on where and how the section is measured. This work
presented a workflow that assesses the whole channel from 3D scans of a
master and its replica, without sectioning. We developed open-source
software that implements it.

The workflow characterises a single channel and compares surface pairs along the fabrication chain: an insert against its design, a part against its insert, and an insert against its initial state over a moulding run. In a comparison, \RQd{}, the correlation of the two slope maps,
tells whether the shape of the channel was reproduced. The extracted width,
depth and wall angle tell whether its size was, and the height and slope
error maps show where a deviation lies.

These outputs can support practical decisions. Ranking insert resins by
\RQd{}, read against the level that an insert reaches against its own
rescan, identifies the resin that replicates the channel best. Following
\RQd{} over a moulding run shows when an insert starts to wear; here it was
the first indicator to move, at shot $5$. Comparing the insert with its
design, and the part with its insert, shows whether a deviation should be
corrected at the printing or at the moulding step; for the smallest
channels it arose in printing.

The evidence supports these uses. On synthetic channels with known ground
truth, \RQd{} fell monotonically as the sidewalls were rounded, more steeply
than the height correlation. It was invariant to uniform amplitude scaling
and carried a registration bias below $0.02$. Across fifteen insert-to-part pairs in three SLA resins, it ranged from $0.17$ to
$0.91$. The height correlation placed the ten well-formed channels within
$0.03$ of each other and ordered one pair against the independently
extracted sidewall geometry. Established image-similarity indices behaved
like the height correlation. For these channels \RQd{} was governed by the
sidewalls. Over a twenty-shot run it separated an insert that settled in the
first shot and then held, flat at $0.88$ against its as-printed state, from a
degrading one ($0.90$ to $0.72$) whose height correlation stayed above
$0.93$; the wall angle tied the drift to wear of the sidewalls and recorded
the settling, which width and depth did not.

The workflow requires only registered areal topographies of a master and its
replica, and it can in principle be applied to other mould-based processes.
Its values are defined at a stated grid cell size, here
\SI{5}{\micro\meter}, and are comparable only at equal cell size. It also
requires registration to within about one grid cell. The results were
obtained on straight channels, with one insert per condition and without a
formal evaluation of measurement uncertainty. Curved and free-form features,
replicate inserts and repeated scans are the next steps.

\printcredits

\section*{Data and code availability}
The point clouds analysed in this work and the Microchannel and Replication
Quality Analyzer (MaRQ, version 1.0), open-source software, are archived
at\newline
\mbox{\url{https://doi.org/10.5281/zenodo.22813916}}. Additional data are
available from the corresponding authors on reasonable request.

\section*{Funding}
This work was funded by TÜBİTAK under grant number 22AG008 within the scope of
the MAESTRO project. This work was also funded by the European Union's Horizon
Europe Programme under grant agreement N°~101079473 within the scope of the
OrChESTRA project.

\section*{Declaration of competing interest}
The authors declare that they have no known competing financial interests or
personal relationships that could have appeared to influence the work reported
in this paper.

\section*{Declaration of generative AI and AI-assisted technologies in the manuscript preparation process}
The authors used AI-assisted technologies in order to assist with editing the
manuscript text. After using this tool, the authors reviewed and edited the
content as needed and take full responsibility for the content of the published
article.

\bibliographystyle{elsarticle-num}
\bibliography{references}

\end{document}


\begin{center}
{\large\bfseries Supplementary material}\\[1ex]
{\bfseries A 3D measurement workflow for assessing replication quality and
tool degradation in moulded microfluidic channels}\\[1ex]
Mehmet Oğulcan Güngör, Jérôme Charmet, Ender Yıldırım
\end{center}

\noindent This supplement gives the algorithm reference of the analysis
software (Section~\ref{sec:s1}), the complete synthetic characterisation of
\RQd{} (Section~\ref{sec:s2}), and the per-channel numbers behind the figures
and tables of the main text (Sections~\ref{sec:s3} to \ref{sec:s9}). All
values were computed with the production code of the software at the fixed
settings of the main text: analysis grid cell size
$\Delta = \SI{5}{\micro\meter}$, the $\sigma = \Delta$ pre-gradient filter
(the software's floor, its ``no smoothing'' setting), channel-interior mask
at $0.20$ of the depth on both surfaces. Only the bare (uncoated) inserts are reported. Section and table
numbers of the form ``Section~3.2'' refer to the main text. HT and R10K
denote the High Temp and Rigid 10K resins.

\section{Algorithm reference of the analysis software}\label{sec:s1}

\begin{figure}[!htbp]
\centering
\includegraphics[width=\textwidth]{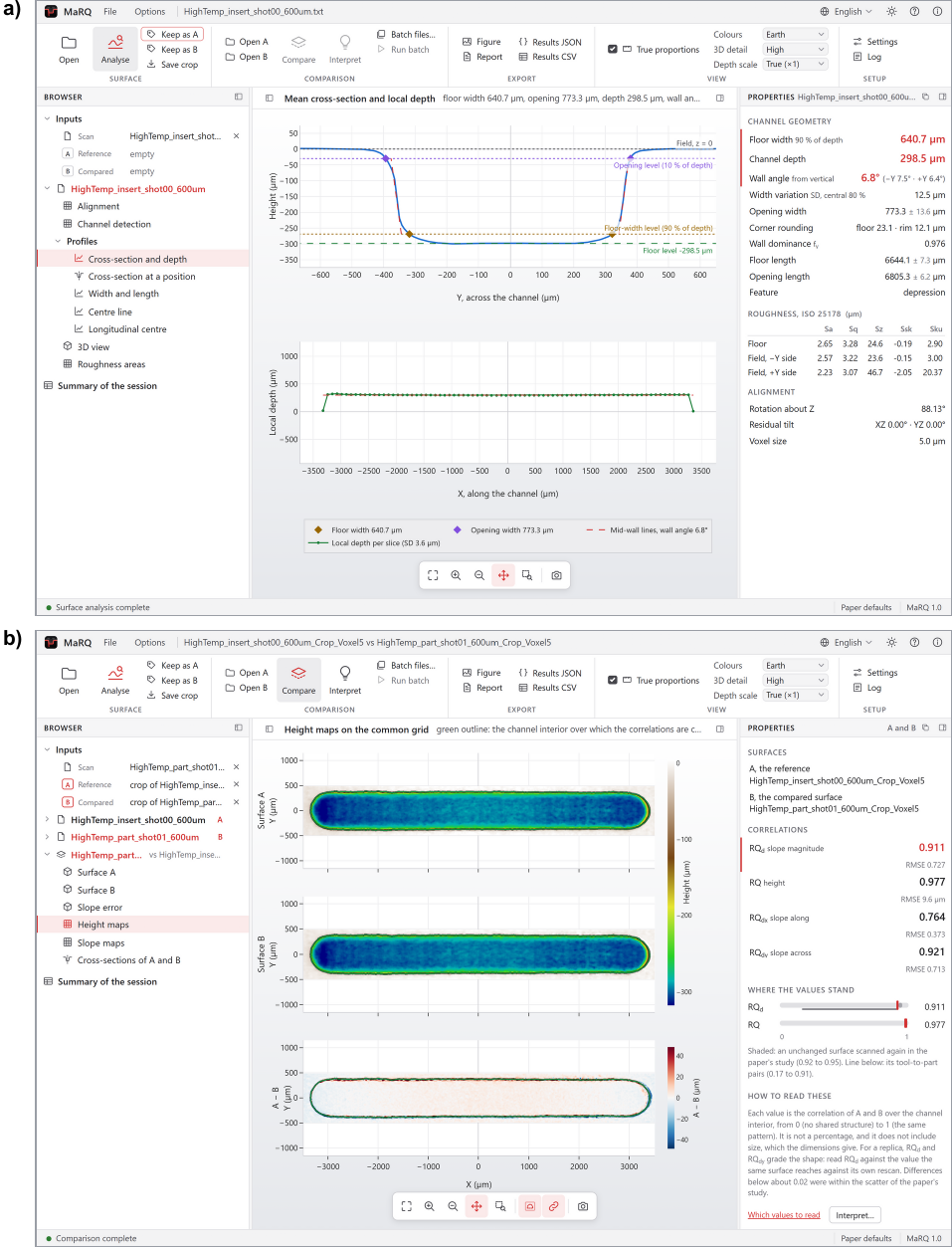}
\caption{The software in a web browser, on the sample channel of the data
package. (a)~Single-surface analysis of a High Temp \SI{600}{\micro\meter}
insert: the commands (top), the inputs and views of each result (left), the
extracted dimensions and roughness parameters (right), and the mean
cross-section and the local depth along the channel.
(b)~Paired comparison of the insert (A) with its first moulded part (B): the
height maps of both surfaces and the height error $E_Z$ on the common grid,
with \RQd{}, \RQ{}, \RQdx{} and \RQdy{}.}
\label{fig:s1_ui}
\end{figure}

The Microchannel and Replication Quality Analyzer (MaRQ, version 1.0) is
implemented in Python and runs client-side in the browser through Pyodide, a WebAssembly build of
the scientific Python stack. It uses NumPy for array and gradient
operations, SciPy (\texttt{scipy.ndimage}, \texttt{scipy.stats}) for Gaussian
filtering, nearest-valid gap filling and the Pearson correlation, Plotly
for the interactive plots and three.js for the 3D views. It accepts delimited text point clouds
(\texttt{.txt}, \texttt{.xyz}, \texttt{.csv}) with automatic delimiter
detection and ISO~25178-72 X3P binary files, whose invalid points, flagged
in a validity mask or stored as missing values, are left out.
The same functions drive a batch runner, so interactive and large-scale runs
execute identical code. Figure~\ref{fig:s1_ui} shows the interface.

\subsection*{Single-surface module}

The single-surface module characterises the geometry and areal roughness of
one surface through the following pipeline.
\begin{enumerate}[wide, itemsep=2pt, topsep=3pt, parsep=0pt]
  \item \textbf{Load} the point cloud and normalise the coordinates to
    micrometres.
  \item \textbf{Edge crop} a fixed margin to discard unreliable edge points.
  \item \textbf{Despike.} A field plane is fitted to an \SI{8}{\micro\meter}
    voxel copy of the cloud. A raised feature, such as an insert ridge, reaches
    further above this plane than below it (by the 99th and 1st percentiles of
    the copy) and is first turned into the equivalent depression by a
    $180^{\circ}$ rotation about an in-plane axis. Because rims rising higher
    than a channel is deep would make it read as a ridge, MaRQ lets the user
    set the feature to channel or ridge instead, and every scan here was left
    on the automatic decision. The points more than
    \SI{40}{\micro\meter} above the plane (focus-variation spikes) are then
    removed, and so are the points far below the channel floor (optical
    dropouts): more than \SI{380}{\micro\meter} below the plane, or $1.15$
    times the floor depth when that is deeper. The floor depth here is that of
    the deepest \SI{5}{\micro\meter} level, among those more than
    \SI{40}{\micro\meter} below the plane, that holds at least a fifth of the
    points of the fullest of them.
  \item \textbf{Voxel downsample} to a uniform point density at
    \SI{5}{\micro\meter}, which regularises the sampling and sets the
    sampling interval of every later quantity.
  \item \textbf{Field levelling and orientation.} The field is fitted by a
    robust least-squares plane: the inliers are seeded on the outer border
    of the scan, which is field, and re-selected in six passes as the points
    within $2.5$ median absolute deviations of the plane. The plane is
    subtracted, which places the field at $z = 0$. The orientation is checked
    again on the levelled heights, by their 2nd and 98th percentiles, and the
    field is re-zeroed.
  \item \textbf{Channel-axis alignment.} Principal component analysis of the
    points below half the estimated depth gives the channel axis, which is
    rotated onto $x$ with the channel centred at the origin; the field plane
    is then fitted again in this frame and subtracted, leaving a residual
    tilt below $0.001^{\circ}$.
  \item \textbf{Depth from the aggregate cross-section.} The channel extent
    is the along-axis span over which the channel is deeper than half the
    estimated depth. All points over this extent are pooled into one
    transverse profile of $400$ equally spaced bins, each holding the median
    height of its points (bins with fewer than three points omitted). The
    floor level $z_f$ is the median of the bins at or below the fifth
    percentile, and the channel depth is $d = -z_f$. When the channel is narrow
    against the scan, spanning at half depth less than $6\,\%$ of the profile,
    the floor level is read within twice that width around it, here and in
    every slice.
  \item \textbf{Widths, wall angle and corner rounding on grid lines.} The
    points are averaged on a grid of \SI{5}{\micro\meter} cells across the
    channel, in lines \SI{10}{\micro\meter} apart along it, over the central
    $80\,\%$ of the channel extent, so that the rounded channel ends are
    excluded. On every line the wall position at the level $z = -f\,d$ is the
    first place where the height profile rises above the level, moving
    outwards from the channel centre line on each side, located by linear
    interpolation between the two grid points that bracket it; no crossing is
    interpolated across more than three empty cells. The floor, mid-depth and
    opening widths $w_\mathrm{floor}$, $w_\mathrm{mid}$ and $w_\mathrm{open}$
    are the distances between the two walls at $f = 0.90$, $0.50$ and $0.10$,
    each reported as the median over the lines. On each line and wall a
    straight line is fitted by least squares to the wall positions at
    $f = 0.25$, $0.375$, $0.50$, $0.625$ and $0.75$; its inclination from the
    vertical, positive when the channel widens towards the opening, is the
    wall angle of that wall. The reported wall angle is the median over the
    lines and both walls; the angles of the two walls are also given. Walls of
    $1^{\circ}$ and $2^{\circ}$ read $1.6^{\circ}$ and $2.4^{\circ}$, so a wall
    within $3^{\circ}$ of the vertical is reported as $<3^{\circ}$. The floor-corner
    rounding is the distance by which the wall at $f = 0.90$ lies inside the
    extended fitted line, and the rim-corner rounding the distance by which
    the wall at $f = 0.10$ lies outside it, both medians over the lines and
    walls.
  \item \textbf{Channel length and crop.} The length is read at $f = 0.90$ in
    the same way on lines along the channel, within a tenth of the floor width
    of its centre line. Each end is found from the outside in, as the
    outermost run of at least three cells below the level and the
    interpolated crossing just outside it, so that a floor that undulates
    across the level does not end the length early. The crop passed to the
    paired-surface module spans the two measured ends plus
    \SI{180}{\micro\meter} along the channel and the opening width plus
    \SI{100}{\micro\meter} on each side across it. On a broken channel the
    crop therefore covers the part that still reaches the floor level, and
    the missing part lies outside the comparison.
  \item \textbf{Longitudinal variation.} The standard deviations of the
    widths over the grid lines describe their variation along the channel.
    For the depth, the channel extent is divided into 80 non-overlapping
    slices. In each slice the transverse points are assigned to 120 equally
    spaced bins and the median height in each bin forms the slice profile;
    bins with fewer than two points are omitted rather than interpolated, and
    slices with fewer than twenty points are discarded. Each accepted slice
    gives a local depth by the floor-level rule above, and the standard
    deviation over the central $80\,\%$ of the valid slices, the first and
    last tenth in order along the axis being excluded, is reported. These
    spreads describe variation within one scan, not measurement
    repeatability, and are not an uncertainty of the reported values.
  \item \textbf{Roughness.} ISO~25178-2 areal parameters are computed on
    three regions. The floor strip is $0.5$ times the nominal channel width
    wide and \SI{4000}{\micro\meter} long and takes the points deeper than
    $0.7\,d$. The two field strips are $400 \times \SI{4000}{\micro\meter}$,
    take the points above $-0.3\,d$, and are placed automatically beyond the
    channel rim detected from the transverse profile, symmetric about the
    channel axis; the left and right strips are reported separately. Each
    region is levelled by subtracting its own least-squares plane, and the
    parameters are evaluated on the residual points at the
    \SI{5}{\micro\meter} sampling, with no further short- or long-wavelength
    filtering.
  \item \textbf{Wall dominance.} The cropped points are averaged on a grid of
    the voxel size, each cell centred on its point, smoothed by the one-cell
    NaN-aware Gaussian filter of the paired-surface module and differentiated
    by central differences. Over the cells deeper than $0.20$ of the depth,
    taken here as the 1st percentile of the smoothed heights, within the
    central $60\,\%$ of the crop length, the wall-dominance ratio
    $f_y = \sum_k (\partial Z/\partial y)_k^2 / \sum_k |\nabla Z|_k^2$ is the
    share of the squared slope magnitude carried by the slope across the
    channel. The rounded channel ends, where the slope runs along the channel,
    are left out.
\end{enumerate}

\noindent On the height residuals $z_i$ ($i = 1, \dots, N$) of a region after
plane removal, the reported parameters are
\begin{gather}
S_a = \frac{1}{N}\sum_{i=1}^{N} |z_i|, \qquad
S_q = \sqrt{\frac{1}{N}\sum_{i=1}^{N} z_i^{2}}, \qquad
S_z = \max_i z_i - \min_i z_i, \\
S_{sk} = \frac{1}{N S_q^{3}}\sum_{i=1}^{N} z_i^{3}, \qquad
S_{ku} = \frac{1}{N S_q^{4}}\sum_{i=1}^{N} z_i^{4}.
\end{gather}
Every step is deterministic with the fixed parameters of
Table~\ref{tab:s1_params}; there is no random subsampling, randomised
threshold or Monte Carlo iteration, so repeated analyses of the same scan
return identical numbers. The batch runner used for this study, which calls
the same functions, additionally flags implausible results for inspection:
a residual tilt above $0.1^{\circ}$, a depth above \SI{340}{\micro\meter}, an
opening narrower than the floor, an inclination between the floor and
opening wall positions outside $-2^{\circ}$ to $35^{\circ}$, or a floor width
outside $0.5$ to $2.5$ times the nominal.

\begin{table}[!htbp]
\centering
\caption{Parameters of the single-surface module, fixed for every result in
the paper.}
\label{tab:s1_params}
\small
\begin{tabularx}{\textwidth}{P{0.23\textwidth} P{0.31\textwidth} X}
\toprule
Parameter & Value & Role \\
\midrule
Edge margin & \SI{100}{\micro\meter} & Scan border discarded before processing \\
Orientation & 99th against 1st percentile before the despike; 98th against 2nd after levelling & A raised feature turned over ($180^{\circ}$ about an in-plane axis) \\
Despiking & above $+40$; below $-380$, or $-1.15\,d_f$ when deeper (\si{\micro\meter}) & Field-plane residual limits; $d_f$ the floor depth found on the \SI{8}{\micro\meter} copy \\
Voxel size & \SI{5}{\micro\meter} & Downsampling cell size and sampling interval \\
Field plane inliers & within 2.5 median absolute deviations, 6 passes & Robust least-squares datum, seeded on the scan border \\
Residual field tilt & $< 0.001^{\circ}$ & After re-levelling in the aligned frame \\
Display grids & square cells, automatic size, not finer than the voxel & Visualisation only \\
Aggregate profile & 400 bins, median per bin, $\geq 3$ points & All points over the channel extent pooled; depth \\
Floor level & median of bins $\le$ 5th percentile & Floor level $z_f$ of a profile; within twice the half-depth width of a channel narrower than 6\,\% of the profile \\
Channel depth & $d = -z_f$ & Field at $z = 0$ after levelling \\
Grid lines & \SI{5}{\micro\meter} cells across, lines \SI{10}{\micro\meter} apart, central 80\,\% of the extent & Widths, wall angle and corner rounding; medians over the lines \\
Wall crossing & linear between the bracketing grid points, $\leq 3$ empty cells between them & Wall position at a level \\
Width levels & 0.90 / 0.50 / 0.10 of the depth & $w_\mathrm{floor}$, $w_\mathrm{mid}$, $w_\mathrm{open}$ \\
Wall angle & least-squares line through the wall at 0.25 to 0.75 of the depth & From the vertical, per wall; $<3^{\circ}$ not resolved \\
Corner rounding & offset of the wall from that line at 0.90 and 0.10 of the depth & Floor and rim corners \\
Length & lines within $0.1\,w_\mathrm{floor}$ of the centre line; ends from the outside in, $\geq 3$ cells & At 0.90 of the depth \\
Crop & measured ends $+$ \SI{180}{\micro\meter}; opening width $+ 2 \times$ \SI{100}{\micro\meter} & Input of the paired-surface module \\
Longitudinal slices & 80 & Non-overlapping, linearly spaced along the channel axis; depth \\
Transverse profile bins & 120 & Bins across the scan width per slice \\
Bin height statistic & median & Height assigned to each transverse bin \\
Minimum points per bin & 2 & Bins with fewer points are omitted, not interpolated \\
Minimum points per slice & 20 & Slices with fewer points are omitted \\
Longitudinal spread & SD over the grid lines; depth: over the central 80\,\% of the slices & First and last tenth of the valid slices excluded \\
Floor roughness strip & $0.5\times$ nominal width $\times$ \SI{4000}{\micro\meter} & Centred on the floor, points deeper than $0.7\,d$ \\
Field roughness strips & $400 \times \SI{4000}{\micro\meter}$ & Symmetric about the axis, beyond the detected rim, points above $-0.3\,d$ \\
Roughness form removal & least-squares plane & Per region; no further filtering \\
Wall dominance $f_y$ & central 60\,\% of the crop, cells below $0.2\,d$ & Share of the squared slope across the channel \\
\bottomrule
\end{tabularx}
\end{table}

\subsection*{Paired-surface module}

The paired-surface module quantifies the similarity of two surfaces, each
already aligned to the canonical channel frame by the single-surface module.
The two clouds are resampled onto a common grid spanning their overlapping
extent by bin averaging, at the correlation grid cell size
$\Delta = \SI{5}{\micro\meter}$, cells without a point being empty; a
NaN-aware Gaussian filter of $\sigma = \max(\sigma_{\mathrm{user}}, \Delta)$
is applied, that is $\sigma = \Delta$ for the paper's setting
$\sigma_{\mathrm{user}} = 0$, which regularises the field and fills an
empty cell from measured cells up to four cells away; a common valid-cell mask
excludes any cell still missing on either surface; and the height gradients
are computed by second-order central differences over $2\Delta$. Because the
clouds were thinned to \SI{5}{\micro\meter} voxels by the single-surface
module, about $7\,\%$ of the cells are empty before the filter (below
$0.2\,\%$ with native-density clouds); this raises \RQd{} by less than
$0.01$ on average and by at most $0.02$ (main text, Section~2.7). Because a straight channel is nearly symmetric about its axis,
the relative in-plane orientation of a pair is ambiguous by $180^{\circ}$;
the module evaluates both orientations and keeps the one with the higher
\RQd{} (\RQ{} against an idealised CAD surface), a correction required for 12
of the 15 insert-to-part comparisons of the paper. The correlations are then
evaluated over the channel interior, the cells lying below $0.20$ of the
depth on both surfaces, with the depth taken as the mean of the 1st
percentiles of the two height maps on the common grid. With $Z_A$ and $Z_B$
the co-registered height maps and $\partial Z / \partial x$,
$\partial Z / \partial y$ the gradients along and across the channel,
\begin{gather}
\RQ = \mathrm{Pearson}(Z_A, Z_B), \\
\RQdx = \mathrm{Pearson}\!\left(\frac{\partial Z_A}{\partial x}, \frac{\partial Z_B}{\partial x}\right), \qquad
\RQdy = \mathrm{Pearson}\!\left(\frac{\partial Z_A}{\partial y}, \frac{\partial Z_B}{\partial y}\right), \\
\RQd = \mathrm{Pearson}\!\left(|\nabla Z_A|, |\nabla Z_B|\right), \qquad
|\nabla Z| = \sqrt{\left(\frac{\partial Z}{\partial x}\right)^{2} + \left(\frac{\partial Z}{\partial y}\right)^{2}},
\end{gather}
where for two fields $a$ and $b$ sampled over the $N$ retained cells
\begin{equation}
\mathrm{Pearson}(a, b) =
\frac{\sum_{i=1}^{N}(a_i - \bar{a})(b_i - \bar{b})}
     {\sqrt{\sum_{i=1}^{N}(a_i - \bar{a})^{2}}\,\sqrt{\sum_{i=1}^{N}(b_i - \bar{b})^{2}}},
\qquad \bar{a} = \frac{1}{N}\sum_{i=1}^{N} a_i .
\end{equation}
Each metric is reported with its root-mean-square difference, and two
spatially resolved error maps retain the absolute information the
correlations discard: the signed height difference $E_Z = Z_A - Z_B$ and the
unsigned gradient-magnitude difference
$E_d = \bigl|\,|\nabla Z_B| - |\nabla Z_A|\,\bigr|$. When one surface is the
idealised CAD reference, recognised by a floor carrying less than
\SI{1.5}{\micro\meter} of texture, the height correlation does not saturate
and the difference is dominated by systematic wall offsets, which the
extracted dimensions measure directly. The software then gives the height
correlation as the key result, with the gradient correlations alongside, and
the paper reports the height correlation and the dimensions for these pairs.

\subsection*{Coordinate conventions}

After alignment the $x$ axis runs along the channel, the $y$ axis across it,
and $z$ is the surface height; the channel is centred at the origin with the
field at $z = 0$. Because the channel is aligned with the axes, the $x$
component of the gradient describes variation along the channel and the $y$
component cuts through the sidewalls.

\subsection*{Validation of the geometry extraction}

The geometry readings were checked against surfaces of known shape at the
settings of the paper. Table~\ref{tab:s1_cad} compares the five CAD
references with their STL design files, read on the same levels and lines
as the software reads them: the widths agree within \SI{0.1}{\micro\meter}
at all three levels, the wall angle within $0.02^{\circ}$ of the designed
$10^{\circ}$, and the floor-level length within \SI{5}{\micro\meter}.
Table~\ref{tab:s1_synthetic} takes synthetic channels of known geometry
through the complete single-surface module. Walls of $5^{\circ}$ to
$30^{\circ}$ are recovered within $0.3^{\circ}$ and the widths within a few
micrometres, also under correlated texture, random wall dropout of up to
$75\,\%$ of the points, and misalignment. Wall dropout in patches costs up to
$0.4^{\circ}$. Two limits follow from the definitions.
Walls of $1^{\circ}$ and $2^{\circ}$ read $1.6^{\circ}$ and $2.4^{\circ}$, so a
wall within $3^{\circ}$ of the vertical is reported as $<3^{\circ}$. A floor fillet that
reaches into the $25$ to $75\,\%$ band leaves less straight wall to fit, and
the wall angle then includes part of the fillet. For circular fillets on
$5^{\circ}$ to $15^{\circ}$ walls the bias stays below $0.5^{\circ}$ while the
measured floor-corner rounding is below \SI{20}{\micro\meter}, and grows to
about $1^{\circ}$ at \SI{25}{\micro\meter} and $2^{\circ}$ to $3^{\circ}$ at
$30$ to \SI{33}{\micro\meter} ($+3^{\circ}$ and $+8.5^{\circ}$ for 150 and
\SI{200}{\micro\meter} fillets on a \SI{600}{\micro\meter} channel). The
measured floor-corner rounding reaches $28$ to \SI{33}{\micro\meter} on a few
as-printed inserts (Table~\ref{tab:s4_dimensions}), whose wall angles may
therefore read up to about $2^{\circ}$ high if their corners are circular. In
both wall-angle trends of the main text, the first-shot change of the High
Temp inserts and the drift of the Rigid 10K inserts, the floor-corner rounding
falls while the wall angle grows, so this bias reduces the reported increases
rather than producing them. Table~\ref{tab:s1_settings} changes one grid-line
setting at a time on the measured scans, and Table~\ref{tab:s1_precision}
reads every scan on two disjoint halves of its grid lines; both effects are
far below the differences discussed in the main text. The two walls of a
channel differ consistently: the left wall of the Rigid 10K inserts leans
about $3^{\circ}$ further from the vertical than the right, mirrored in their
parts, while the two CAD walls agree within $0.04^{\circ}$.

\begin{table}[!htbp]
\centering
\caption{The five CAD channels against their design: dimensions read from the STL files on the same levels and lines as the software reads them (widths at 90, 50 and 10\,\% of the \SI{300}{\micro\meter} depth across the middle of the channel, length at the floor level on the centre line, $10^{\circ}$ walls), and the error of the extracted value (extracted $-$ design), in \si{\micro\meter} unless stated.}
\label{tab:s1_cad}
\small
\begin{tabular}{@{}rccccccccc@{}}
\toprule
Nominal & \multicolumn{2}{c}{Floor width} & \multicolumn{2}{c}{Mid-depth width} & \multicolumn{2}{c}{Opening width} & \multicolumn{2}{c}{Length} & Wall angle \\
\cmidrule(lr){2-3}\cmidrule(lr){4-5}\cmidrule(lr){6-7}\cmidrule(lr){8-9}\cmidrule(lr){10-10}
(\si{\micro\meter}) & design & error & design & error & design & error & design & error & error ($^{\circ}$) \\
\midrule
100 & 110.6 & $-$0.03 & 152.9 & 0.10 & 195.2 & $-$0.02 & 6107 & 0.7 & $-$0.02 \\
200 & 210.6 & $-$0.05 & 252.9 & 0.06 & 295.2 & $-$0.02 & 6209 & $-$4.3 & $-$0.01 \\
300 & 310.6 & $-$0.06 & 352.9 & 0.07 & 395.2 & $-$0.06 & 6304 & $-$0.1 & $-$0.02 \\
600 & 610.6 & $-$0.08 & 652.9 & 0.08 & 695.2 & 0.07 & 6601 & $-$0.4 & 0.00 \\
1000 & 1010.6 & $-$0.07 & 1052.9 & 0.10 & 1095.2 & $-$0.01 & 7001 & $-$0.6 & $-$0.01 \\
\bottomrule
\end{tabular}
\end{table}

\begin{table}[!htbp]
\centering
\caption{Synthetic channels of known geometry (widths $100$ to \SI{1000}{\micro\meter}, \SI{300}{\micro\meter} deep, rounded ends, sampled at the instrument's \SI{3.51}{\micro\meter} pitch at random sub-pitch offsets) taken through the complete single-surface module at the paper's settings: mean error and, in brackets, largest absolute error (extracted $-$ true), in \si{\micro\meter} unless stated. Fillets are circular floor and rim fillets on 100 and \SI{600}{\micro\meter} channels with $5$ to $15^{\circ}$ walls; channels whose floor fillet is wider than the floor are excluded (no flat floor). Texture is Gaussian-correlated over \SI{12}{\micro\meter}.}
\label{tab:s1_synthetic}
\footnotesize
\setlength{\tabcolsep}{4pt}
\begin{tabular}{@{}p{0.25\textwidth}rcccccc@{}}
\toprule
Synthetic channels & Cases & Floor & Mid-depth & Opening & Wall & Floor & Length \\
 & & width & width & width & angle ($^{\circ}$) & corner & \\
\midrule
Sharp corners, walls $5$--$30^{\circ}$ & 90 & $-$0.5 (3.2) & 0.2 (1.2) & 0.0 (1.6) & $-$0.02 (0.26) & 0.3 (2.2) & $-$2.5 (6.8) \\
Sharp corners, walls $1$--$3^{\circ}$ & 45 & $-$2.5 (4.5) & 0.0 (2.2) & 2.8 (5.8) & 0.37 (0.91) & 0.5 (1.3) & $-$5.1 (9.3) \\
Fillets $\leq$ \SI{200}{\micro\meter}, clear of the 25--75\,\% band & 44 & $-$0.4 (2.1) & $-$0.2 (1.0) & 0.4 (2.0) & $-$0.04 (0.25) & 0.4 (1.3) & $-$2.7 (5.1) \\
Fillets reaching into the 25--75\,\% band & 16 & $-$0.4 (2.7) & $-$0.2 (1.1) & 0.2 (1.3) & 4.35 (11.74) & $-$13.0 (35.3) & $-$3.2 (4.6) \\
Texture, SD $1$--\SI{5}{\micro\meter} & 60 & $-$0.9 (2.2) & 0.2 (1.0) & 0.1 (0.3) & $-$0.14 (0.25) & 0.9 (1.6) & $-$2.2 (10.0) \\
Random wall dropout $10$--$75\,\%$ & 24 & $-$1.1 (1.9) & 0.1 (0.9) & 0.2 (0.5) & $-$0.09 (0.23) & 0.8 (1.3) & $-$2.8 (6.2) \\
Random wall dropout $90\,\%$ & 6 & $-$2.8 (3.4) & 0.3 (0.8) & 1.3 (1.5) & $-$0.06 (0.14) & 1.4 (1.7) & $-$6.0 (9.9) \\
Wall dropout in patches, $10$--$50\,\%$ & 18 & $-$0.6 (1.3) & 0.0 (1.0) & 0.1 (0.2) & 0.03 (0.41) & 0.1 (1.5) & $-$2.1 (8.8) \\
Rotation $0.5$--$15^{\circ}$, tilt $0$--$3^{\circ}$ & 28 & $-$0.5 (1.4) & 0.2 (0.7) & 0.1 (0.2) & 0.07 (0.25) & 0.0 (1.2) & $-$2.2 (4.4) \\
\bottomrule
\end{tabular}
\end{table}

\begin{table}[!htbp]
\centering
\caption{Measured scans with one grid-line setting changed from its default (lines \SI{10}{\micro\meter} apart, gap limit 3 cells, central 80\,\%): largest absolute change over the CAD, as-printed insert and first-part scans of High Temp and Rigid 10K (35 surfaces), in \si{\micro\meter} unless stated, and of $\mathrm{RQ}_{d}$ over their ten insert-to-part pairs, recomputed on the crops of each setting.}
\label{tab:s1_settings}
\small
\setlength{\tabcolsep}{4pt}
\begin{tabular}{@{}lcccccc@{}}
\toprule
Setting changed & Floor width & Opening width & Wall angle ($^{\circ}$) & Rim corner & Length & $\mathrm{RQ}_{d}$ \\
\midrule
Line spacing \SI{5}{\micro\meter} & 0.4 & 0.5 & 0.14 & 0.3 & 1.9 & 0.0010 \\
Line spacing \SI{20}{\micro\meter} & 0.5 & 0.6 & 0.15 & 0.5 & 2.3 & 0.0022 \\
Gap limit 1 cell & 0.0 & 0.1 & 0.00 & 0.0 & 1.4 & 0.0000 \\
Gap limit 6 cells & 0.0 & 0.0 & 0.00 & 0.0 & 0.1 & 0.0000 \\
Central 60\,\% of the length & 3.8 & 3.5 & 0.44 & 2.0 & 1.0 & 0.0016 \\
Central 90\,\% of the length & 1.7 & 4.9 & 0.25 & 0.8 & 0.0 & 0.0027 \\
\bottomrule
\end{tabular}
\end{table}

\begin{table}[!htbp]
\centering
\caption{Dependence of the reported medians on the grid lines used: each scan read on its even and on its odd grid lines only, two disjoint halves; largest half-difference over the scans of the group (CAD, as-printed and first-shot inserts and first parts), in \si{\micro\meter} unless stated. This is the sampling component of the extraction, not a repeatability of the measurement.}
\label{tab:s1_precision}
\small
\begin{tabular}{@{}p{0.2\textwidth}rccccccc@{}}
\toprule
Scans & $n$ & Floor & Mid-depth & Opening & Wall & Floor & Rim & Length \\
 & & width & width & width & angle ($^{\circ}$) & corner & corner & \\
\midrule
CAD references & 5 & 0.04 & 0.06 & 0.05 & 0.01 & 0.05 & 0.04 & 0.6 \\
HT and R10K & 30 & 0.61 & 0.22 & 0.66 & 0.06 & 0.38 & 0.25 & 2.2 \\
White & 15 & 1.49 & 0.61 & 0.81 & 0.06 & 0.22 & 0.06 & 6.5 \\
\bottomrule
\end{tabular}
\end{table}

\section{Synthetic characterisation of \texorpdfstring{\RQd}{RQd} in full}\label{sec:s2}

The synthetic surfaces are those of Section~2.4 of the main text: a
trapezoidal channel \SI{300}{\micro\meter} deep with a \SI{600}{\micro\meter}
floor and a $10^{\circ}$ draft on a $4 \times 2$~mm field sampled at
\SI{5}{\micro\meter}, with a band-limited random microtexture of
\SI{3}{\micro\meter} standard deviation and about \SI{12}{\micro\meter}
correlation length. A degraded copy is compared with the undegraded
reference through the production pipeline. The degradation operators are:
wall smoothing (Gaussian rounding of the cross-section profile of scale
$\sigma$, texture intact); global smoothing (isotropic Gaussian blur of the
whole textured field); depth fill (depth reduced at fixed opening width);
positive amplitude scaling of the whole height field; additive Gaussian
noise; a single localised Gaussian feature on the channel rim; and a textured
field with the channel removed. Rigid translations and rotations are imposed
on the point coordinates and passed through the production registration.

\subsection*{Smoothing sweeps}

Table~\ref{tab:s2_smoothing} gives the full sweeps of Fig.~5(e,f) of the
main text with the across-channel component added. \RQdy{} falls with
\RQd{} throughout, which is the synthetic form of the wall dominance measured
on the real channels. Wall and global smoothing agree within $0.015$ up to
\SI{100}{\micro\meter}; above it the global blur, which also removes the
microtexture, decays more slowly, because the intact texture of the
wall-smoothed surface competes with its weakened walls in the gradient map.

\begin{table}[!htbp]
\centering
\caption{Smoothing sweeps on the synthetic channel: Gaussian rounding of the cross-section profile (wall smoothing) and isotropic Gaussian blur of the whole textured field (global smoothing), both of scale $\sigma$.}
\label{tab:s2_smoothing}
\small
\begin{tabular}{@{}rccccc@{}}
\toprule
$\sigma$ (\si{\micro\meter}) & \multicolumn{3}{c}{Wall smoothing} & \multicolumn{2}{c}{Global smoothing} \\
\cmidrule(lr){2-4}\cmidrule(lr){5-6}
 & $\mathrm{RQ}$ & $\mathrm{RQ}_{d}$ & $\mathrm{RQ}_{dy}$ & $\mathrm{RQ}_{d}$ & $\mathrm{RQ}$ \\
\midrule
0 & 1.000 & 1.000 & 1.000 & 1.000 & 1.000 \\
10 & 0.998 & 0.993 & 0.994 & 0.992 & 0.998 \\
20 & 0.983 & 0.970 & 0.971 & 0.965 & 0.982 \\
30 & 0.951 & 0.926 & 0.929 & 0.917 & 0.950 \\
40 & 0.911 & 0.872 & 0.878 & 0.860 & 0.909 \\
50 & 0.870 & 0.818 & 0.829 & 0.805 & 0.868 \\
60 & 0.831 & 0.768 & 0.784 & 0.754 & 0.830 \\
80 & 0.764 & 0.683 & 0.710 & 0.671 & 0.763 \\
100 & 0.712 & 0.612 & 0.651 & 0.609 & 0.711 \\
120 & 0.672 & 0.551 & 0.605 & 0.560 & 0.671 \\
140 & 0.643 & 0.497 & 0.568 & 0.521 & 0.642 \\
160 & 0.623 & 0.451 & 0.541 & 0.490 & 0.622 \\
180 & 0.611 & 0.413 & 0.523 & 0.466 & 0.610 \\
200 & 0.604 & 0.384 & 0.511 & 0.452 & 0.603 \\
\bottomrule
\end{tabular}
\end{table}

\subsection*{Depth fill, noise and amplitude scaling}

Table~\ref{tab:s2_other} gives the depth-fill and noise sweeps of Fig.~5(g)
with the height correlation alongside, and the amplitude-scaling check.
Both correlations are blind to a uniform loss of depth, because both are
amplitude-invariant, and the depth must be read from the extracted dimensions.
Their fall at the largest losses comes from the channel-interior mask, which
shrinks once the filled channel is barely deeper than the mask level (to about
$6\,\%$ of its cells at $90\,\%$). Under positive amplitude scaling both
correlations stay exactly at $1$ while the mean gradient error grows with
the scale factor, which confirms that absolute-amplitude information resides
in the error maps and dimensions, not in the correlations.

\begin{table}[!htbp]
\centering
\caption{Depth fill at fixed opening width, added Gaussian noise, and positive amplitude scaling of the whole height field.}
\label{tab:s2_other}
\small
\begin{tabular}{@{}lrcc@{}}
\toprule
Degradation & Amount & $\mathrm{RQ}_{d}$ & $\mathrm{RQ}$ \\
\midrule
Depth lost (\%) & 0 & 1.000 & 1.000 \\
Depth lost (\%) & 10 & 1.000 & 1.000 \\
Depth lost (\%) & 20 & 1.000 & 1.000 \\
Depth lost (\%) & 30 & 1.000 & 1.000 \\
Depth lost (\%) & 40 & 0.999 & 0.999 \\
Depth lost (\%) & 50 & 0.998 & 0.998 \\
Depth lost (\%) & 60 & 0.994 & 0.993 \\
Depth lost (\%) & 65 & 0.991 & 0.987 \\
Depth lost (\%) & 70 & 0.984 & 0.975 \\
Depth lost (\%) & 75 & 0.970 & 0.946 \\
Depth lost (\%) & 80 & 0.932 & 0.853 \\
Depth lost (\%) & 85 & 0.786 & 0.619 \\
Depth lost (\%) & 90 & 0.425 & 0.495 \\
\addlinespace[2pt]
Noise SD (\si{\micro\meter}) & 0 & 1.000 & 1.000 \\
Noise SD (\si{\micro\meter}) & 5 & 0.997 & 1.000 \\
Noise SD (\si{\micro\meter}) & 10 & 0.989 & 0.998 \\
Noise SD (\si{\micro\meter}) & 20 & 0.957 & 0.993 \\
Noise SD (\si{\micro\meter}) & 30 & 0.900 & 0.984 \\
Noise SD (\si{\micro\meter}) & 40 & 0.822 & 0.971 \\
Noise SD (\si{\micro\meter}) & 50 & 0.731 & 0.955 \\
Noise SD (\si{\micro\meter}) & 60 & 0.637 & 0.935 \\
Noise SD (\si{\micro\meter}) & 70 & 0.548 & 0.913 \\
Noise SD (\si{\micro\meter}) & 80 & 0.470 & 0.888 \\
\addlinespace[2pt]
Amplitude factor & 0.5 & 1.000 & 1.000 \\
Amplitude factor & 0.8 & 1.000 & 1.000 \\
Amplitude factor & 1.25 & 1.000 & 1.000 \\
Amplitude factor & 2 & 1.000 & 1.000 \\
Amplitude factor & 5 & 1.000 & 1.000 \\
\bottomrule
\end{tabular}
\end{table}

\subsection*{Gradient-energy decomposition}

On the reference surface the sidewalls occupy about $5\,\%$ of the analysed
cells but carry about $94\,\%$ of the squared gradient magnitude over the
whole field and about $97\,\%$ within the channel-interior mask; the floor
contributes $2$ to $3\,\%$ and the field the remainder. The geometry-only
wall energy exceeds the textured-floor energy by a factor of about 39. This
is the origin of the wall dominance that makes \RQd{} track \RQdy{} for
these channels.

\subsection*{Seed robustness}

Table~\ref{tab:s2_seeds} repeats selected severities over 25 microtexture
seeds, the reference and the degraded surface sharing the seed. The spread
is far below the differences between adjacent severity levels, so the
monotonic trends are not single-seed artefacts. These intervals describe the
sensitivity to the artificial texture realisation, not experimental
uncertainty. The flat-field anchor of the main text, two fields with the
channel removed and independent textures, is a separate test over 30 seeds
and returns $-0.001 \pm 0.009$.

\begin{table}[!htbp]
\centering
\caption{Seed robustness: $\mathrm{RQ}_{d}$ over 25 microtexture seeds (reference and degraded surface share the seed).}
\label{tab:s2_seeds}
\small
\begin{tabular}{@{}lrcccc@{}}
\toprule
Degradation & Amount & Mean & SD & Min & Max \\
\midrule
Noise SD (\si{\micro\meter}) & 20 & 0.956 & 0.000 & 0.955 & 0.958 \\
Noise SD (\si{\micro\meter}) & 40 & 0.819 & 0.002 & 0.816 & 0.823 \\
Wall smoothing $\sigma$ (\si{\micro\meter}) & 40 & 0.871 & 0.002 & 0.868 & 0.875 \\
Wall smoothing $\sigma$ (\si{\micro\meter}) & 80 & 0.682 & 0.004 & 0.677 & 0.691 \\
Wall smoothing $\sigma$ (\si{\micro\meter}) & 200 & 0.385 & 0.010 & 0.370 & 0.404 \\
\bottomrule
\end{tabular}
\end{table}

\subsection*{Registration bias}

Uncorrected rigid misalignment strongly reduces \RQd{}: an across-channel
shift of \SI{20}{\micro\meter} gives $0.72$ and one of \SI{50}{\micro\meter}
a slightly negative value; a $1^{\circ}$ rotation gives $0.77$ and
$2^{\circ}$ gives $0.45$. When both surfaces are passed through the
production registration the imposed transformations are removed almost
completely: translations and rotations both recover to $\RQd = 1.000$. Table~\ref{tab:s2_registration} summarises the residual bias over
19 imposed transforms (translations up to \SI{100}{\micro\meter} along and
across the channel, rotations of $0.1^{\circ}$ to $2^{\circ}$) for five
degradation states. The bias is at most $0.002$,
with no registration failure; a small rotation residual of the severe case, largest at
$2^{\circ}$, is the largest term. The severe-case baseline here ($0.385$)
differs slightly from the $0.384$ of the canonical sweep because the
registration test routes each surface through the full point-cloud path
(voxel downsampling, cropping and re-gridding) whereas the canonical sweep
evaluates the metric directly on the generated grid.

\begin{table}[!htbp]
\centering
\caption{Post-registration bias of \RQd{} over 19 imposed rigid transforms.}
\label{tab:s2_registration}
\small
\begin{tabular}{@{}lcccc@{}}
\toprule
Degradation state & Aligned baseline & Mean bias & Max $|$bias$|$ & $|$bias$| < 0.02$ \\
\midrule
identical            & 1.000 & $0.000$ & 0.000 & 19/19 \\
moderate ($\sigma = 80$)  & 0.683 & $0.000$ & 0.000 & 19/19 \\
severe ($\sigma = 200$)   & 0.385 & $0.000$ & 0.002 & 19/19 \\
global smoothing     & 0.754 & $0.000$ & 0.000 & 19/19 \\
independent texture  & 0.996 & $0.000$ & 0.000 & 19/19 \\
\bottomrule
\end{tabular}
\end{table}

\subsection*{Localised-defect detection limit}

A single Gaussian feature was swept in amplitude and width. Even the largest
tested, \SI{250}{\micro\meter} in amplitude and \SI{400}{\micro\meter} wide,
changed \RQd{} by less than $0.003$, and none moved it by $0.01$, whereas
the maximum of $E_d$ scaled with the amplitude. Localised damage occupying
a small fraction of the area is diluted in the global scalar and must be
read from the error maps, which is the division of labour used in
Section~3.3 of the main text.

\subsection*{Channel width and normalisation}

For channels of $100$, $200$, $300$, $600$ and \SI{1000}{\micro\meter} floor
width at fixed depth and draft, the wall-smoothing curves nearly collapse
against the absolute smoothing scale but not against the scale normalised by
floor width: the mean standard deviation across widths is $0.008$ for the
absolute scale and $0.192$ for the width-normalised scale. Because the walls
are geometrically identical across widths, a given absolute smoothing costs
a similar \RQd{} regardless of floor width, even though it removes a larger
fraction of a narrow channel; proportional severity must be read from the
dimensional measurements.

\subsection*{Directional components}

In synthetic cases with a shared texture, across-channel defects load onto
\RQdy{} while \RQdx{} stays near unity, and along-channel or floor
perturbations load onto \RQdx{}. In the experimental comparisons the same
wall dominance appears: \RQdy{} tracks \RQd{} with a correlation of $0.97$,
while \RQdx{} is systematically low and remains low even for the same insert
measured before and after a shot, because fine floor texture does not
reproduce between independent measurements. \RQdx{} therefore carries a
relative trend but its absolute value is not a floor-fidelity score.

\subsection*{Equivalent smoothing scales}

Because the wall-smoothing response is monotonic, a measured \RQd{} can be
read back as an equivalent Gaussian smoothing scale of the synthetic channel.
For the first-shot resin means of the main text this gives about
\SI{\eqHT}{\micro\meter} for High Temp ($\RQd = 0.87$) and about
\SI{\eqRTK}{\micro\meter} for Rigid~10K ($0.81$); the White value ($0.28$)
lies below the range of the pure wall-smoothing sweep, indicating a more
severe loss of correspondence or additional deformation modes. These are
metric-equivalent severities under the chosen synthetic geometry and
settings, not physical rounding radii.

\section{The fifteen insert-to-part comparisons}\label{sec:s3}

Table~\ref{tab:s3_pairs} lists every pair behind Fig.~6 and Table~3 of the
main text.

\begin{table}[!htbp]
\centering
\caption{The fifteen insert-to-part comparisons (first moulded part against its as-printed insert, channel-interior mask, 5~\si{\micro\meter} grid): the four correlations, the wall-dominance ratio $f_y$, the fraction of grid cells valid on both surfaces and the mean absolute height error.}
\label{tab:s3_pairs}
\small
\begin{tabular}{@{}lrccccccc@{}}
\toprule
Resin & Width & $\mathrm{RQ}$ & $\mathrm{RQ}_{dx}$ & $\mathrm{RQ}_{dy}$ & $\mathrm{RQ}_{d}$ & $f_y$ & Valid & $\overline{|E_Z|}$ \\
 & (\si{\micro\meter}) & & & & & & (\%) & (\si{\micro\meter}) \\
\midrule
High Temp & 100 & 0.957 & 0.560 & 0.867 & 0.815 & 0.982 & 100 & 7.5 \\
High Temp & 200 & 0.966 & 0.614 & 0.906 & 0.880 & 0.984 & 100 & 5.7 \\
High Temp & 300 & 0.975 & 0.587 & 0.919 & 0.895 & 0.983 & 100 & 4.7 \\
High Temp & 600 & 0.977 & 0.764 & 0.921 & 0.911 & 0.983 & 100 & 4.5 \\
High Temp & 1000 & 0.970 & 0.747 & 0.881 & 0.870 & 0.982 & 100 & 5.9 \\
Rigid 10K & 100 & 0.980 & 0.577 & 0.903 & 0.825 & 0.990 & 100 & 11.0 \\
Rigid 10K & 200 & 0.990 & 0.679 & 0.926 & 0.883 & 0.989 & 100 & 9.5 \\
Rigid 10K & 300 & 0.982 & 0.594 & 0.888 & 0.843 & 0.988 & 100 & 10.3 \\
Rigid 10K & 600 & 0.972 & 0.592 & 0.779 & 0.723 & 0.974 & 100 & 11.3 \\
Rigid 10K & 1000 & 0.983 & 0.688 & 0.821 & 0.779 & 0.962 & 100 & 10.7 \\
White & 100 & 0.535 & 0.133 & 0.581 & 0.390 & 0.969 & 71 & 60.8 \\
White & 200 & 0.453 & 0.182 & 0.328 & 0.173 & 0.922 & 66 & 66.1 \\
White & 300 & 0.467 & -0.041 & 0.458 & 0.304 & 0.931 & 68 & 64.8 \\
White & 600 & 0.428 & 0.244 & 0.229 & 0.258 & 0.892 & 82 & 40.9 \\
White & 1000 & 0.435 & 0.284 & 0.203 & 0.266 & 0.864 & 88 & 28.0 \\
\bottomrule
\end{tabular}
\end{table}

\section{Channel geometry through the chain}\label{sec:s4}

Table~\ref{tab:s4_dimensions} extends the floor widths and depths plotted in Fig.~9
of the main text to the opening width, the wall angle and the corner
rounding of every CAD reference, as-printed insert and first moulded part.

\begin{table}[!htbp]
\centering
\caption{Extracted channel geometry through the fabrication chain for the CAD reference, the as-printed insert and its first moulded part: floor and opening widths, depth, and floor- and rim-corner rounding in \si{\micro\meter}, wall angle in degrees from the vertical ($<3$: within $3^{\circ}$ of the vertical, below the resolution of the grid lines).}
\label{tab:s4_dimensions}
\small
\begin{tabular}{@{}rlccccccc@{}}
\toprule
Nom. & Quantity & CAD & \multicolumn{2}{c}{High Temp} & \multicolumn{2}{c}{Rigid 10K} & \multicolumn{2}{c}{White} \\
\cmidrule(lr){4-5}\cmidrule(lr){6-7}\cmidrule(lr){8-9}
(\si{\micro\meter}) & & & Insert & Part & Insert & Part & Insert & Part \\
\midrule
100 & floor width & 111 & 195 & 200 & 171 & 166 & 213 & 192 \\
 & opening width & 195 & 312 & 307 & 315 & 338 & 273 & 211 \\
 & depth & 300 & 293 & 294 & 274 & 269 & 311 & 309 \\
 & wall angle ($^{\circ}$) & 10.0 & 6.3 & 5.7 & 14.3 & 12.8 & $<3$ & $<3$ \\
 & floor corner & 0 & 17 & 22 & 14 & 14 & 14 & 1 \\
 & rim corner & 0 & 14 & 4 & 5 & 26 & 0 & 0 \\
\addlinespace[2pt]
200 & floor width & 211 & 263 & 266 & 241 & 236 & 269 & 212 \\
 & opening width & 295 & 391 & 390 & 396 & 426 & 319 & 228 \\
 & depth & 300 & 302 & 303 & 291 & 287 & 310 & 307 \\
 & wall angle ($^{\circ}$) & 10.0 & 6.3 & 5.8 & 12.8 & 12.6 & $<3$ & $<3$ \\
 & floor corner & 0 & 23 & 27 & 22 & 19 & 13 & 0 \\
 & rim corner & 0 & 11 & 7 & 4 & 24 & 0 & 0 \\
\addlinespace[2pt]
300 & floor width & 311 & 337 & 341 & 317 & 311 & 354 & 276 \\
 & opening width & 395 & 475 & 477 & 478 & 515 & 397 & 289 \\
 & depth & 300 & 302 & 303 & 295 & 290 & 308 & 301 \\
 & wall angle ($^{\circ}$) & 10.0 & 6.6 & 6.7 & 10.8 & 11.5 & $<3$ & $<3$ \\
 & floor corner & 0 & 27 & 28 & 30 & 26 & 10 & 0 \\
 & rim corner & 0 & 10 & 7 & 6 & 31 & 0 & 0 \\
\addlinespace[2pt]
600 & floor width & 610 & 641 & 646 & 612 & 606 & 656 & 585 \\
 & opening width & 695 & 773 & 772 & 757 & 814 & 689 & 601 \\
 & depth & 300 & 299 & 300 & 297 & 291 & 301 & 298 \\
 & wall angle ($^{\circ}$) & 10.0 & 6.8 & 6.4 & 10.3 & 11.8 & $<3$ & $<3$ \\
 & floor corner & 0 & 23 & 24 & 26 & 20 & 4 & 0 \\
 & rim corner & 0 & 12 & 9 & 4 & 35 & 0 & 0 \\
\addlinespace[2pt]
1000 & floor width & 1011 & 1047 & 1050 & 985 & 981 & 1056 & 981 \\
 & opening width & 1095 & 1175 & 1177 & 1156 & 1223 & 1088 & 1002 \\
 & depth & 300 & 298 & 302 & 303 & 296 & 297 & 300 \\
 & wall angle ($^{\circ}$) & 10.0 & 9.9 & 8.4 & 11.4 & 13.4 & $<3$ & $<3$ \\
 & floor corner & 0 & 19 & 23 & 32 & 25 & 4 & 0 \\
 & rim corner & 0 & 2 & 3 & 3 & 34 & 0 & 0 \\
\bottomrule
\end{tabular}
\end{table}

\section{Areal roughness by channel}\label{sec:s5}

Table~\ref{tab:s5_roughness} gives the per-channel roughness behind Table~4
of the main text.

\begin{table}[!htbp]
\centering
\caption{Areal roughness (\si{\micro\meter}, ISO 25178-2) of the channel floor of every as-printed insert and its first moulded part, and of the insert field (mean of the two field strips).}
\label{tab:s5_roughness}
\small
\begin{tabular}{@{}lrccccccc@{}}
\toprule
Resin & Width & \multicolumn{3}{c}{Insert floor} & \multicolumn{3}{c}{Part floor} & Insert field \\
\cmidrule(lr){3-5}\cmidrule(lr){6-8}\cmidrule(lr){9-9}
 & (\si{\micro\meter}) & $S_a$ & $S_q$ & $S_z$ & $S_a$ & $S_q$ & $S_z$ & $S_a$ \\
\midrule
High Temp & 100 & 3.08 & 3.80 & 23.0 & 2.74 & 3.42 & 20.9 & 2.51 \\
High Temp & 200 & 2.85 & 3.42 & 19.6 & 2.55 & 3.11 & 18.6 & 2.33 \\
High Temp & 300 & 2.61 & 3.22 & 19.2 & 2.34 & 2.92 & 23.7 & 2.29 \\
High Temp & 600 & 2.63 & 3.25 & 24.6 & 2.27 & 2.82 & 21.5 & 2.40 \\
High Temp & 1000 & 2.29 & 2.81 & 21.1 & 1.85 & 2.30 & 18.1 & 2.10 \\
Rigid 10K & 100 & 3.82 & 4.66 & 25.6 & 3.47 & 4.24 & 23.8 & 4.77 \\
Rigid 10K & 200 & 2.64 & 3.29 & 28.0 & 2.91 & 3.53 & 19.7 & 4.13 \\
Rigid 10K & 300 & 2.62 & 3.27 & 28.4 & 2.73 & 3.34 & 23.0 & 4.45 \\
Rigid 10K & 600 & 2.93 & 3.70 & 27.6 & 3.22 & 3.88 & 21.1 & 4.12 \\
Rigid 10K & 1000 & 4.59 & 5.93 & 41.0 & 4.45 & 5.58 & 38.1 & 4.44 \\
White & 100 & 3.62 & 4.40 & 24.9 & 3.36 & 4.05 & 21.8 & 3.65 \\
White & 200 & 3.69 & 4.53 & 26.4 & 3.42 & 4.23 & 28.3 & 3.36 \\
White & 300 & 2.99 & 3.67 & 22.4 & 3.12 & 4.06 & 29.3 & 3.37 \\
White & 600 & 2.81 & 3.51 & 25.4 & 2.78 & 3.51 & 31.9 & 3.15 \\
White & 1000 & 3.01 & 3.73 & 29.3 & 3.82 & 5.51 & 110.4 & 3.23 \\
\bottomrule
\end{tabular}
\end{table}

\section{Insert degradation by channel width}\label{sec:s6}

Tables~\ref{tab:s6_lifetime_HT} and \ref{tab:s6_lifetime_R10K} give, for
every width and shot, the numbers averaged in Fig.~8 of the main text. The
White insert reached end of life within five shots and is omitted.

{\renewcommand{\arraystretch}{0.92}\begin{table}[!htbp]
\centering
\caption{High Temp insert over the moulding run, by channel width: correlations between the shot-0 scan and the scan at shot $N$, and the geometry and floor roughness of each scan.}
\label{tab:s6_lifetime_HT}
\small
\begin{tabular}{@{}rrccccccccc@{}}
\toprule
Width & Shot & $\mathrm{RQ}_{d}$ & $\mathrm{RQ}_{dy}$ & $\mathrm{RQ}$ & Wall & Floor & Depth & $S_a$ & $S_z$ \\
(\si{\micro\meter}) & & \multicolumn{3}{c}{shot 0 $\to$ shot $N$} & ($^{\circ}$) & (\si{\micro\meter}) & (\si{\micro\meter}) & (\si{\micro\meter}) & (\si{\micro\meter}) \\
\midrule
100 & 0 & -- & -- & -- & 6.3 & 195 & 293 & 3.08 & 23 \\
 & 1 & 0.860 & 0.903 & 0.970 & 11.7 & 196 & 294 & 3.12 & 21 \\
 & 5 & 0.870 & 0.912 & 0.972 & 13.2 & 192 & 293 & 3.15 & 21 \\
 & 10 & 0.867 & 0.909 & 0.970 & 13.6 & 194 & 292 & 3.12 & 22 \\
 & 20 & 0.856 & 0.902 & 0.960 & 11.8 & 197 & 292 & 3.14 & 22 \\
200 & 0 & -- & -- & -- & 6.3 & 263 & 302 & 2.85 & 20 \\
 & 1 & 0.875 & 0.906 & 0.973 & 12.5 & 264 & 303 & 2.86 & 18 \\
 & 5 & 0.862 & 0.896 & 0.966 & 13.7 & 261 & 301 & 2.98 & 19 \\
 & 10 & 0.875 & 0.905 & 0.974 & 12.8 & 261 & 300 & 2.98 & 19 \\
 & 20 & 0.874 & 0.906 & 0.968 & 11.7 & 259 & 301 & 2.93 & 20 \\
300 & 0 & -- & -- & -- & 6.6 & 337 & 302 & 2.61 & 19 \\
 & 1 & 0.877 & 0.903 & 0.974 & 13.0 & 339 & 303 & 2.61 & 20 \\
 & 5 & 0.862 & 0.878 & 0.949 & 13.1 & 339 & 301 & 2.70 & 32 \\
 & 10 & 0.879 & 0.905 & 0.975 & 12.7 & 337 & 295 & 2.71 & 20 \\
 & 20 & 0.885 & 0.911 & 0.974 & 11.6 & 337 & 300 & 2.63 & 22 \\
600 & 0 & -- & -- & -- & 6.8 & 641 & 299 & 2.63 & 25 \\
 & 1 & 0.896 & 0.911 & 0.979 & 13.3 & 645 & 299 & 2.57 & 23 \\
 & 5 & 0.883 & 0.899 & 0.977 & 13.8 & 644 & 296 & 2.61 & 23 \\
 & 10 & 0.881 & 0.897 & 0.975 & 14.2 & 642 & 289 & 2.65 & 24 \\
 & 20 & 0.896 & 0.911 & 0.977 & 13.1 & 641 & 290 & 2.82 & 53 \\
1000 & 0 & -- & -- & -- & 9.9 & 1047 & 298 & 2.29 & 21 \\
 & 1 & 0.902 & 0.917 & 0.980 & 12.8 & 1049 & 299 & 2.21 & 20 \\
 & 5 & 0.882 & 0.898 & 0.976 & 14.7 & 1048 & 298 & 2.30 & 21 \\
 & 10 & 0.886 & 0.898 & 0.980 & 12.8 & 1046 & 296 & 2.26 & 20 \\
 & 20 & 0.893 & 0.905 & 0.981 & 12.1 & 1045 & 296 & 2.31 & 21 \\
\bottomrule
\end{tabular}
\end{table}
}
{\renewcommand{\arraystretch}{0.92}\begin{table}[!htbp]
\centering
\caption{Rigid 10K insert over the moulding run, by channel width: correlations between the shot-0 scan and the scan at shot $N$, and the geometry and floor roughness of each scan.}
\label{tab:s6_lifetime_R10K}
\small
\begin{tabular}{@{}rrccccccccc@{}}
\toprule
Width & Shot & $\mathrm{RQ}_{d}$ & $\mathrm{RQ}_{dy}$ & $\mathrm{RQ}$ & Wall & Floor & Depth & $S_a$ & $S_z$ \\
(\si{\micro\meter}) & & \multicolumn{3}{c}{shot 0 $\to$ shot $N$} & ($^{\circ}$) & (\si{\micro\meter}) & (\si{\micro\meter}) & (\si{\micro\meter}) & (\si{\micro\meter}) \\
\midrule
100 & 0 & -- & -- & -- & 14.3 & 171 & 274 & 3.82 & 26 \\
 & 1 & 0.890 & 0.937 & 0.983 & 12.2 & 169 & 275 & 3.88 & 27 \\
 & 5 & 0.809 & 0.897 & 0.903 & 14.4 & 156 & 271 & 4.97 & 47 \\
 & 10 & 0.777 & 0.881 & 0.904 & 16.6 & 150 & 264 & 4.09 & 36 \\
 & 20 & 0.724 & 0.855 & 0.885 & 20.4 & 140 & 239 & 9.00 & 59 \\
200 & 0 & -- & -- & -- & 12.8 & 241 & 291 & 2.64 & 28 \\
 & 1 & 0.915 & 0.946 & 0.992 & 11.7 & 240 & 292 & 2.91 & 22 \\
 & 5 & 0.890 & 0.930 & 0.987 & 14.4 & 226 & 289 & 3.80 & 31 \\
 & 10 & 0.836 & 0.897 & 0.969 & 16.7 & 218 & 281 & 3.35 & 42 \\
 & 20 & 0.732 & 0.834 & 0.940 & 19.9 & 178 & 264 & 7.06 & 57 \\
300 & 0 & -- & -- & -- & 10.8 & 317 & 295 & 2.62 & 28 \\
 & 1 & 0.900 & 0.928 & 0.986 & 9.7 & 316 & 296 & 2.76 & 24 \\
 & 5 & 0.859 & 0.900 & 0.980 & 12.9 & 309 & 291 & 3.12 & 29 \\
 & 10 & 0.767 & 0.834 & 0.952 & 14.7 & 300 & 286 & 2.95 & 27 \\
 & 20 & 0.644 & 0.751 & 0.928 & 18.4 & 262 & 283 & 7.29 & 71 \\
600 & 0 & -- & -- & -- & 10.3 & 612 & 297 & 2.93 & 28 \\
 & 1 & 0.868 & 0.891 & 0.982 & 10.4 & 612 & 295 & 3.38 & 23 \\
 & 5 & 0.869 & 0.893 & 0.971 & 12.3 & 603 & 291 & 3.10 & 25 \\
 & 10 & 0.799 & 0.838 & 0.962 & 14.3 & 594 & 287 & 2.70 & 24 \\
 & 20 & 0.679 & 0.744 & 0.945 & 17.5 & 571 & 278 & 2.21 & 29 \\
1000 & 0 & -- & -- & -- & 11.4 & 985 & 303 & 4.59 & 41 \\
 & 1 & 0.919 & 0.932 & 0.990 & 11.5 & 987 & 301 & 4.63 & 41 \\
 & 5 & 0.908 & 0.924 & 0.986 & 12.3 & 978 & 297 & 4.72 & 39 \\
 & 10 & 0.885 & 0.906 & 0.981 & 13.9 & 977 & 294 & 4.42 & 36 \\
 & 20 & 0.818 & 0.854 & 0.968 & 16.8 & 969 & 284 & 3.46 & 31 \\
\bottomrule
\end{tabular}
\end{table}
}

Table~\ref{tab:s6_first_shot} follows the wall shape through the first shots.
In the first shot the High Temp walls straightened and tilted: the wall angle
rose by $3^{\circ}$ to $6.5^{\circ}$, the floor- and rim-corner rounding fell
by $2$ to \SI{14}{\micro\meter} and the opening widened by $5$ to
\SI{12}{\micro\meter}, while the floor width changed by less than
\SI{5}{\micro\meter} and the depth by at most \SI{1}{\micro\meter}. The part
moulded in that shot, scanned in the same session as the changed insert,
keeps the as-printed shape, and the later parts follow the changed insert, so
the change is a change of the insert and not of the measurement. The Rigid 10K
inserts show no such step; their wall angle grows steadily with the drift, and
the rim of their first parts is rounded by about \SI{30}{\micro\meter}, a
rounding the later parts do not show. Table~\ref{tab:s6_same_state} compares
each insert with itself between any two shots. For High Temp every comparison
with shot $0$ reads $0.87$ to $0.88$ and every comparison among the later
states $0.92$ to $0.95$; for Rigid 10K the values fall with the number of
shots between the scans.

\begin{table}[!htbp]
\centering
\caption{Wall shape of the High Temp and Rigid 10K inserts and their parts through the first shots, mean $\pm$ standard deviation over the five channel widths; corner rounding and the width changes, taken per width against Insert 0, in \si{\micro\meter}. Part 1 was moulded in the first shot and scanned with Insert 1; Part 5 was moulded in the fifth shot.}
\label{tab:s6_first_shot}
\small
\setlength{\tabcolsep}{4pt}
\begin{tabular}{@{}llcccccc@{}}
\toprule
Resin & Quantity & Insert 0 & Insert 1 & Part 1 & Insert 5 & Part 5 & Insert 20 \\
\midrule
High Temp & wall angle ($^{\circ}$) & 7.2\,$\pm$\,1.5 & 12.7\,$\pm$\,0.6 & 6.6\,$\pm$\,1.1 & 13.7\,$\pm$\,0.6 & 11.3\,$\pm$\,0.5 & 12.1\,$\pm$\,0.6 \\
 & floor corner & 22\,$\pm$\,4 & 13\,$\pm$\,3 & 25\,$\pm$\,2 & 10\,$\pm$\,5 & 17\,$\pm$\,2 & 16\,$\pm$\,5 \\
 & rim corner & 10\,$\pm$\,4 & 0\,$\pm$\,0 & 6\,$\pm$\,3 & 0\,$\pm$\,0 & 1\,$\pm$\,0 & 0\,$\pm$\,0 \\
 & floor width change & 0\,$\pm$\,0 & 2\,$\pm$\,1 & 4\,$\pm$\,1 & 1\,$\pm$\,3 & 3\,$\pm$\,1 & 0\,$\pm$\,2 \\
 & opening width change & 0\,$\pm$\,0 & 8\,$\pm$\,3 & $-$1\,$\pm$\,3 & 12\,$\pm$\,2 & 5\,$\pm$\,3 & 4\,$\pm$\,2 \\
\addlinespace[2pt]
Rigid 10K & wall angle ($^{\circ}$) & 11.9\,$\pm$\,1.6 & 11.1\,$\pm$\,1.0 & 12.4\,$\pm$\,0.8 & 13.3\,$\pm$\,1.1 & 14.3\,$\pm$\,1.0 & 18.6\,$\pm$\,1.5 \\
 & floor corner & 25\,$\pm$\,7 & 25\,$\pm$\,6 & 21\,$\pm$\,5 & 23\,$\pm$\,6 & 19\,$\pm$\,6 & 17\,$\pm$\,4 \\
 & rim corner & 4\,$\pm$\,1 & 7\,$\pm$\,2 & 30\,$\pm$\,5 & 5\,$\pm$\,1 & 7\,$\pm$\,5 & 6\,$\pm$\,2 \\
 & floor width change & 0\,$\pm$\,0 & 0\,$\pm$\,2 & $-$5\,$\pm$\,1 & $-$11\,$\pm$\,4 & $-$15\,$\pm$\,4 & $-$41\,$\pm$\,19 \\
 & opening width change & 0\,$\pm$\,0 & 1\,$\pm$\,3 & 43\,$\pm$\,19 & 0\,$\pm$\,4 & $-$1\,$\pm$\,10 & $-$1\,$\pm$\,4 \\
\bottomrule
\end{tabular}
\end{table}

\begin{table}[!htbp]
\centering
\caption{Each insert compared with itself: $\mathrm{RQ}_{d}$ between its scans after two different shots, mean over the five widths. The first row of each resin is the tool-degradation trajectory of Fig.~8 of the main text.}
\label{tab:s6_same_state}
\small
\begin{tabular}{@{}lrcccc@{}}
\toprule
Resin & From shot & \multicolumn{4}{c}{To shot} \\
\cmidrule(lr){3-6}
 & & 1 & 5 & 10 & 20 \\
\midrule
High Temp & 0 & 0.882 & 0.872 & 0.877 & 0.881 \\
 & 1 &  & 0.942 & 0.938 & 0.925 \\
 & 5 &  &  & 0.947 & 0.924 \\
 & 10 &  &  &  & 0.930 \\
\addlinespace[2pt]
Rigid 10K & 0 & 0.898 & 0.867 & 0.813 & 0.720 \\
 & 1 &  & 0.907 & 0.858 & 0.749 \\
 & 5 &  &  & 0.914 & 0.815 \\
 & 10 &  &  &  & 0.890 \\
\bottomrule
\end{tabular}
\end{table}

\section{Benchmark per pair}\label{sec:s7}

Tables~\ref{tab:s7_bench_pairs} to \ref{tab:s7_bench_drift} give the
individual values summarised in Table~5 of the main text. The White pairs
show the unbounded behaviour of the NIST difference: their part surfaces
differ from the insert by more than the insert's own relief inside the mask,
so $D_s$ exceeds one and $1 - D_s$ becomes strongly negative.

The pointwise gradient-magnitude similarity of Xue et al.\ (reference 35 of
the main text) is evaluated at every cell $k$ of the channel-interior mask as
\begin{equation}
  \mathrm{GMS}_k = \frac{2\,G_{A,k}\,G_{B,k} + c}{G_{A,k}^2 + G_{B,k}^2 + c},
\end{equation}
where $G_{A,k}$ and $G_{B,k}$ are the gradient magnitudes $|\nabla Z|$ of the
two surfaces, from the same central differences as $\mathrm{RQ}_d$, and
$c = 0.0026$ on heights scaled jointly to $[0, 1]$. GMSD is the standard
deviation of $\mathrm{GMS}_k$ over the mask and GMSM its mean.

\begin{table}[!htbp]
\centering
\caption{The six similarity measures of the benchmark for each of the fifteen insert-to-part pairs, evaluated on the same registered grids and channel-interior masks.}
\label{tab:s7_bench_pairs}
\small
\begin{tabular}{@{}lrcccccc@{}}
\toprule
Resin & Width (\si{\micro\meter}) & $\mathrm{RQ}$ & $\mathrm{RQ}_{d}$ & $1 - D_s$ & SSIM & GMSD & GMSM \\
\midrule
High Temp & 100 & 0.957 & 0.815 & 0.90 & 0.949 & 0.081 & 0.969 \\
High Temp & 200 & 0.966 & 0.880 & 0.93 & 0.969 & 0.061 & 0.982 \\
High Temp & 300 & 0.975 & 0.895 & 0.95 & 0.978 & 0.048 & 0.988 \\
High Temp & 600 & 0.977 & 0.911 & 0.95 & 0.985 & 0.035 & 0.992 \\
High Temp & 1000 & 0.970 & 0.870 & 0.94 & 0.985 & 0.039 & 0.992 \\
Rigid 10K & 100 & 0.980 & 0.825 & 0.92 & 0.946 & 0.058 & 0.975 \\
Rigid 10K & 200 & 0.990 & 0.883 & 0.96 & 0.971 & 0.041 & 0.986 \\
Rigid 10K & 300 & 0.982 & 0.843 & 0.93 & 0.967 & 0.056 & 0.983 \\
Rigid 10K & 600 & 0.972 & 0.723 & 0.89 & 0.963 & 0.074 & 0.981 \\
Rigid 10K & 1000 & 0.983 & 0.779 & 0.94 & 0.970 & 0.053 & 0.988 \\
White & 100 & 0.535 & 0.390 & $-$6.45 & 0.745 & 0.270 & 0.880 \\
White & 200 & 0.453 & 0.173 & $-$7.94 & 0.748 & 0.276 & 0.876 \\
White & 300 & 0.467 & 0.304 & $-$42.74 & 0.775 & 0.272 & 0.892 \\
White & 600 & 0.428 & 0.258 & $-$9.63 & 0.872 & 0.193 & 0.947 \\
White & 1000 & 0.435 & 0.266 & $-$7.05 & 0.902 & 0.154 & 0.963 \\
\bottomrule
\end{tabular}
\end{table}

\begin{table}[!htbp]
\centering
\caption{The six measures along the synthetic wall-smoothing series.}
\label{tab:s7_bench_synthetic}
\small
\begin{tabular}{@{}rcccccc@{}}
\toprule
$\sigma$ (\si{\micro\meter}) & $\mathrm{RQ}$ & $\mathrm{RQ}_{d}$ & $1 - D_s$ & SSIM & GMSD & GMSM \\
\midrule
0 & 1.000 & 1.000 & 1.00 & 1.000 & 0.000 & 1.000 \\
10 & 0.998 & 0.993 & 1.00 & 0.994 & 0.009 & 0.998 \\
20 & 0.983 & 0.970 & 0.97 & 0.973 & 0.028 & 0.989 \\
40 & 0.911 & 0.872 & 0.82 & 0.916 & 0.063 & 0.966 \\
60 & 0.831 & 0.768 & 0.65 & 0.872 & 0.095 & 0.950 \\
80 & 0.764 & 0.683 & 0.50 & 0.845 & 0.120 & 0.940 \\
120 & 0.672 & 0.551 & 0.30 & 0.816 & 0.152 & 0.930 \\
160 & 0.623 & 0.451 & 0.26 & 0.805 & 0.171 & 0.925 \\
200 & 0.604 & 0.384 & 0.30 & 0.803 & 0.184 & 0.923 \\
\bottomrule
\end{tabular}
\end{table}

\begin{table}[!htbp]
\centering
\caption{The six measures between the shot-0 insert and the insert at shot $N$, mean over the five widths.}
\label{tab:s7_bench_drift}
\small
\begin{tabular}{@{}lrcccccc@{}}
\toprule
Resin & Shot & $\mathrm{RQ}$ & $\mathrm{RQ}_{d}$ & $1 - D_s$ & SSIM & GMSD & GMSM \\
\midrule
High Temp & 1 & 0.975 & 0.882 & 0.945 & 0.972 & 0.047 & 0.984 \\
 & 5 & 0.968 & 0.872 & 0.931 & 0.968 & 0.052 & 0.982 \\
 & 10 & 0.975 & 0.877 & 0.945 & 0.971 & 0.049 & 0.983 \\
 & 20 & 0.972 & 0.881 & 0.938 & 0.969 & 0.048 & 0.984 \\
Rigid 10K & 1 & 0.987 & 0.898 & 0.969 & 0.976 & 0.039 & 0.989 \\
 & 5 & 0.965 & 0.867 & 0.908 & 0.954 & 0.042 & 0.984 \\
 & 10 & 0.953 & 0.813 & 0.856 & 0.932 & 0.051 & 0.978 \\
 & 20 & 0.933 & 0.720 & 0.709 & 0.895 & 0.062 & 0.969 \\
\bottomrule
\end{tabular}
\end{table}

\section{Sensitivity sweeps in full}\label{sec:s8}

Table~\ref{tab:s8_sensitivity} gives all four one-at-a-time sweeps of
Section~3.9 of the main text, and Table~\ref{tab:s8_sensitivity_lifetime}
the grid-spacing sweep on the tool-degradation trajectories.
Tables~\ref{tab:s8_masks} to \ref{tab:s8_masks_synthetic} give the
robustness check of the channel-interior region-of-interest (ROI) definition (Section~3.9):
the fifteen pairs, the tool-degradation trajectories and a synthetic wall
displacement under the default intersection mask and under masks defined
by the reference surface alone, the replica alone, and their union. The
four masks share the production threshold, grids, orientation and
valid-data mask and differ only in the geometric combination; the
intersection values reproduce the production values exactly. For the
failed White channels the reference-only region is 8 to 22\,\% of the
combined footprint and consists of insert walls and floor where the part
surface lies at field level, which is why the reference-defined mask
returns about zero there; for HT and R10K the disagreement is a rim band
one to three cells wide plus channel ends, and its inclusion moves
$\mathrm{RQ}_{d}$ by at most $0.048$.

\begin{table}[!htbp]
\centering
\caption{One-at-a-time sensitivity of the fifteen insert-to-part correlations to the analysis settings (defaults: grid cell size \SI{5}{\micro\meter}, fraction 0.20, smoothing setting 0, shift 0). Spread: range of $\mathrm{RQ}_{d}$ over the ten HT and R10K channels. The \SI{3.5}{\micro\meter} rows are evaluated on clouds thinned to \SI{3}{\micro\meter} voxels, all others on \SI{5}{\micro\meter} voxels; the smoothing settings 0 and 5 both give $\sigma = \SI{5}{\micro\meter}$, the software's floor of one cell.}
\label{tab:s8_sensitivity}
\small
\begin{tabular}{@{}lrccccccc@{}}
\toprule
Setting & Value & \multicolumn{3}{c}{$\mathrm{RQ}_{d}$, mean over widths} & \multicolumn{3}{c}{$\mathrm{RQ}$, mean over widths} & Spread \\
\cmidrule(lr){3-5}\cmidrule(lr){6-8}
 & & HT & R10K & White & HT & R10K & White & HT+R10K \\
\midrule
Grid cell size $\Delta$ (\si{\micro\meter}) & 3.5 & 0.810 & 0.730 & 0.233 & 0.961 & 0.977 & 0.413 & 0.23 \\
 & 5 & 0.874 & 0.811 & 0.278 & 0.969 & 0.982 & 0.464 & 0.19 \\
 & 10 & 0.958 & 0.917 & 0.384 & 0.983 & 0.988 & 0.580 & 0.10 \\
 & 20 & 0.990 & 0.974 & 0.582 & 0.994 & 0.993 & 0.690 & 0.03 \\
\addlinespace[2pt]
Channel-interior fraction & 0 & 0.851 & 0.742 & 0.021 & 0.996 & 0.992 & 0.627 & 0.24 \\
 & 0.1 & 0.850 & 0.757 & 0.242 & 0.979 & 0.982 & 0.474 & 0.22 \\
 & 0.2 & 0.874 & 0.811 & 0.278 & 0.969 & 0.982 & 0.464 & 0.19 \\
 & 0.4 & 0.882 & 0.884 & 0.304 & 0.963 & 0.978 & 0.425 & 0.08 \\
\addlinespace[2pt]
Pre-gradient smoothing $\sigma$ (\si{\micro\meter}) & 0 & 0.874 & 0.811 & 0.278 & 0.969 & 0.982 & 0.464 & 0.19 \\
 & 5 & 0.874 & 0.811 & 0.278 & 0.969 & 0.982 & 0.464 & 0.19 \\
 & 10 & 0.949 & 0.903 & 0.380 & 0.983 & 0.988 & 0.585 & 0.11 \\
\addlinespace[2pt]
Across-channel shift (\si{\micro\meter}) & 0 & 0.874 & 0.811 & 0.278 & 0.969 & 0.982 & 0.464 & 0.19 \\
 & 5 & 0.854 & 0.820 & 0.298 & 0.944 & 0.973 & 0.477 & 0.16 \\
 & 10 & 0.779 & 0.802 & 0.297 & 0.874 & 0.937 & 0.493 & 0.14 \\
 & 20 & 0.506 & 0.665 & 0.339 & 0.658 & 0.796 & 0.514 & 0.29 \\
\bottomrule
\end{tabular}
\end{table}

\begin{table}[!htbp]
\centering
\caption{Tool-degradation $\mathrm{RQ}_{d}$ (shot 0 to shot $N$, mean over widths) at each grid cell size $\Delta$; the \SI{3.5}{\micro\meter} rows are evaluated on clouds thinned to \SI{3}{\micro\meter} voxels, all others on \SI{5}{\micro\meter} voxels.}
\label{tab:s8_sensitivity_lifetime}
\small
\begin{tabular}{@{}lrcccc@{}}
\toprule
Resin & $\Delta$ (\si{\micro\meter}) & shot 1 & shot 5 & shot 10 & shot 20 \\
\midrule
High Temp & 3.5 & 0.824 & 0.814 & 0.817 & 0.820 \\
 & 5 & 0.882 & 0.872 & 0.877 & 0.881 \\
 & 10 & 0.963 & 0.958 & 0.963 & 0.963 \\
 & 20 & 0.992 & 0.990 & 0.992 & 0.991 \\
Rigid 10K & 3.5 & 0.832 & 0.796 & 0.739 & 0.647 \\
 & 5 & 0.898 & 0.867 & 0.813 & 0.720 \\
 & 10 & 0.973 & 0.952 & 0.915 & 0.841 \\
 & 20 & 0.994 & 0.982 & 0.963 & 0.923 \\
\bottomrule
\end{tabular}
\end{table}

\begin{table}[!htbp]
\centering
\caption{$\mathrm{RQ}_{d}$ of the fifteen insert-to-part pairs under four definitions of the channel-interior ROI: the default intersection (both surfaces below the threshold), the reference (insert) mask alone, the replica (part) mask alone, and their union. Everything else is the production setting: the same registered grids and orientation, the $\sigma = \Delta$ filter, the threshold $-0.20\,d$ with $d$ the joint depth estimate, and the valid-data mask; a cell invalid on either surface is excluded under every definition.}
\label{tab:s8_masks}
\small
\begin{tabular}{@{}lrcccc@{}}
\toprule
Resin & Width (\si{\micro\meter}) & Intersection (default) & Reference & Replica & Union \\
\midrule
High Temp & 100 & 0.815 & 0.814 & 0.781 & 0.780 \\
 & 200 & 0.880 & 0.876 & 0.857 & 0.852 \\
 & 300 & 0.895 & 0.884 & 0.888 & 0.878 \\
 & 600 & 0.911 & 0.908 & 0.905 & 0.901 \\
 & 1000 & 0.870 & 0.863 & 0.867 & 0.860 \\
\addlinespace[2pt]
Rigid 10K & 100 & 0.825 & 0.781 & 0.825 & 0.781 \\
 & 200 & 0.883 & 0.856 & 0.882 & 0.856 \\
 & 300 & 0.843 & 0.813 & 0.842 & 0.812 \\
 & 600 & 0.723 & 0.689 & 0.704 & 0.674 \\
 & 1000 & 0.779 & 0.757 & 0.752 & 0.734 \\
\addlinespace[2pt]
White & 100 & 0.390 & -0.006 & 0.380 & -0.005 \\
 & 200 & 0.173 & 0.000 & 0.172 & 0.001 \\
 & 300 & 0.304 & 0.001 & 0.255 & 0.001 \\
 & 600 & 0.258 & 0.043 & 0.257 & 0.044 \\
 & 1000 & 0.266 & 0.096 & 0.269 & 0.099 \\
\bottomrule
\end{tabular}
\par\smallskip\parbox{0.92\linewidth}{\footnotesize Differences from the default: reference: mean $|\Delta|$ 0.018, max $|\Delta|$ 0.045, $\rho_s$ 0.927 over the ten well-formed channels (0.096, 0.396, 0.932 over all fifteen); replica: mean $|\Delta|$ 0.012, max $|\Delta|$ 0.034, $\rho_s$ 0.988 over the ten well-formed channels (0.012, 0.049, 0.986 over all fifteen); union: mean $|\Delta|$ 0.029, max $|\Delta|$ 0.048, $\rho_s$ 0.964 over the ten well-formed channels (0.103, 0.395, 0.943 over all fifteen).}
\end{table}

\begin{table}[!htbp]
\centering
\caption{Tool-degradation $\mathrm{RQ}_{d}$ (shot 0 to shot $N$, mean over the widths) under the four ROI definitions of Table~\ref{tab:s8_masks}.}
\label{tab:s8_masks_lifetime}
\small
\begin{tabular}{@{}lrcccc@{}}
\toprule
Resin & Shot & Intersection (default) & Reference & Replica & Union \\
\midrule
High Temp & 1 & 0.882 & 0.881 & 0.845 & 0.844 \\
 & 5 & 0.872 & 0.872 & 0.815 & 0.815 \\
 & 10 & 0.877 & 0.877 & 0.839 & 0.839 \\
 & 20 & 0.881 & 0.880 & 0.843 & 0.842 \\
\addlinespace[2pt]
Rigid 10K & 1 & 0.898 & 0.886 & 0.898 & 0.886 \\
 & 5 & 0.867 & 0.850 & 0.866 & 0.848 \\
 & 10 & 0.813 & 0.789 & 0.813 & 0.790 \\
 & 20 & 0.720 & 0.690 & 0.720 & 0.690 \\
\bottomrule
\end{tabular}
\end{table}

\begin{table}[!htbp]
\centering
\caption{Synthetic symmetric wall displacement under the four ROI definitions: both walls of the compared channel moved outward (widening) or inward (narrowing) by $\delta$, depth and draft fixed, same microtexture. The last column is the one-sided wall band that the union adds.}
\label{tab:s8_masks_synthetic}
\small
\begin{tabular}{@{}lrccccc@{}}
\toprule
Mode & $\delta$ (\si{\micro\meter}) & Intersection & Reference & Replica & Union & Union cells over intersection (\%) \\
\midrule
widening & 0 & 1.000 & 1.000 & 1.000 & 1.000 & 0.0 \\
 & 5 & 0.982 & 0.982 & 0.983 & 0.983 & 1.5 \\
 & 10 & 0.934 & 0.934 & 0.932 & 0.932 & 2.9 \\
 & 20 & 0.788 & 0.788 & 0.720 & 0.720 & 5.8 \\
 & 40 & 0.427 & 0.427 & 0.184 & 0.184 & 11.7 \\
 & 80 & 0.039 & 0.039 & -0.138 & -0.138 & 23.4 \\
\addlinespace[2pt]
narrowing & 0 & 1.000 & 1.000 & 1.000 & 1.000 & 0.0 \\
 & 5 & 0.982 & 0.983 & 0.982 & 0.983 & 1.5 \\
 & 10 & 0.933 & 0.931 & 0.933 & 0.931 & 3.0 \\
 & 20 & 0.785 & 0.716 & 0.785 & 0.716 & 6.2 \\
 & 40 & 0.426 & 0.166 & 0.426 & 0.166 & 13.2 \\
 & 80 & 0.042 & -0.175 & 0.042 & -0.175 & 30.5 \\
\bottomrule
\end{tabular}
\end{table}

\section{Layer-line peaks}\label{sec:s9}

Table~\ref{tab:s9_layerlines} gives the background-relative peak heights
quoted in Section~3.7 of the main text.

\begin{table}[!htbp]
\centering
\caption{Along-channel power spectrum of the as-printed insert floor and field, averaged over rows and the five widths, relative to its running median over \SI{25}{\micro\meter} of period. The layer line of a build at $30^{\circ}$ is expected at $\Delta x = h/\sin 30^{\circ} = 2h$.}
\label{tab:s9_layerlines}
\small
\begin{tabular}{@{}lrcccccc@{}}
\toprule
Resin & $h$ (\si{\micro\meter}) & \multicolumn{6}{c}{power / local background at period $\Delta x$ (\si{\micro\meter})} \\
\cmidrule(lr){3-8}
 & & 27 & 36 & 50 & 54 & 100 & 200 \\
\midrule
Rigid 10K & 50 & 1.0 & 0.8 & 1.1 & 1.0 & 16.1 & 1.6 \\
White & 50 & 1.3 & 1.0 & 1.2 & 0.9 & 7.4 & 1.4 \\
High Temp & 25 & 2.6 & 2.0 & 1.1 & 2.8 & 2.2 & 1.0 \\
\bottomrule
\end{tabular}
\end{table}